\documentclass[prl,nobalancelastpage,twocolumn,superscriptaddress,nolongbibliography]{revtex4-2}

\usepackage{color,amsthm,amsmath,amsxtra,amsfonts,dsfont,graphicx,bm,amssymb}
\usepackage[colorlinks=true,linkcolor=blue, citecolor=blue, urlcolor=blue, bookmarks]{hyperref}
\usepackage{centernot}
\usepackage[dvipsnames]{xcolor}
\usepackage{graphicx}
\usepackage{tikz}
\usepackage{braket}
\usepackage{multirow}
\usepackage{makecell}
\newcommand{\prlsection}[1]{{\em {#1}---~}}

\begin{document}

\title{Learning mixed quantum states in large-scale experiments}

\author{Matteo Votto}
\email{matteo.votto@lpmmc.cnrs.fr}
\affiliation{Université Grenoble Alpes, CNRS, LPMMC, 38000 Grenoble, France}

\author{Marko Ljubotina}
\affiliation{Technical University of Munich, TUM School of Natural Sciences, Physics Department, James-Franck-Straße 1, 85748 Garching, Germany}
\affiliation{Munich Center for Quantum Science and Technology (MCQST), Schellingstr. 4, 80799 München, Germany}
\affiliation{Institute of Science and Technology Austria (ISTA), Am Campus 1, 3400 Klosterneuburg, Austria}

\author{Cécilia Lancien}
\affiliation{Université Grenoble Alpes, CNRS, Institut Fourier, 38610 Gières, France}

\author{J. Ignacio Cirac}
\affiliation{Max Planck Institute of Quantum Optics, Hans-Kopfermann-Str. 1, Garching 85748, Germany}	
\affiliation{Munich Center for Quantum Science and Technology (MCQST), Schellingstr. 4, 80799 München, Germany}

\author{Peter Zoller}
\affiliation{Institute for Theoretical Physics, University of Innsbruck, Innsbruck A-6020, Austria}
\affiliation{Institute for Quantum Optics and Quantum Information of the Austrian Academy of Sciences,  Innsbruck A-6020, Austria}

\author{Maksym Serbyn}
\affiliation{Institute of Science and Technology Austria (ISTA), Am Campus 1, 3400 Klosterneuburg, Austria}

\author{Lorenzo Piroli}
\affiliation{Dipartimento di Fisica e Astronomia, Università di Bologna and INFN,
Sezione di Bologna, via Irnerio 46, I-40126 Bologna, Italy}

\author{Benoît Vermersch}
\affiliation{Université Grenoble Alpes, CNRS, LPMMC, 38000 Grenoble, France}
\affiliation{Quobly, Grenoble, France}

\begin{abstract}
    We present and test a protocol to learn the matrix-product operator (MPO) representation of an experimentally prepared quantum state. 
    The protocol takes as an input classical shadows corresponding to local randomized measurements, and outputs the tensors of a MPO which maximizes a suitably-defined fidelity with the experimental state. 
    The tensor optimization is carried out sequentially, similarly to the well-known density matrix renormalization group algorithm. 
    Our approach is provably efficient under certain technical conditions which are expected to be met in short-range correlated states and in typical noisy experimental settings. 
    Under the same conditions, we also provide an efficient scheme to estimate fidelities between the learned and the experimental states.
    We experimentally demonstrate our protocol by learning entangled quantum states of up to $N = 96$ qubits in a superconducting quantum processor.
    Our method upgrades classical shadows to large-scale quantum computation and simulation experiments.
\end{abstract}

\maketitle

\begin{figure*}[t]
    \includegraphics[width=0.99\textwidth]{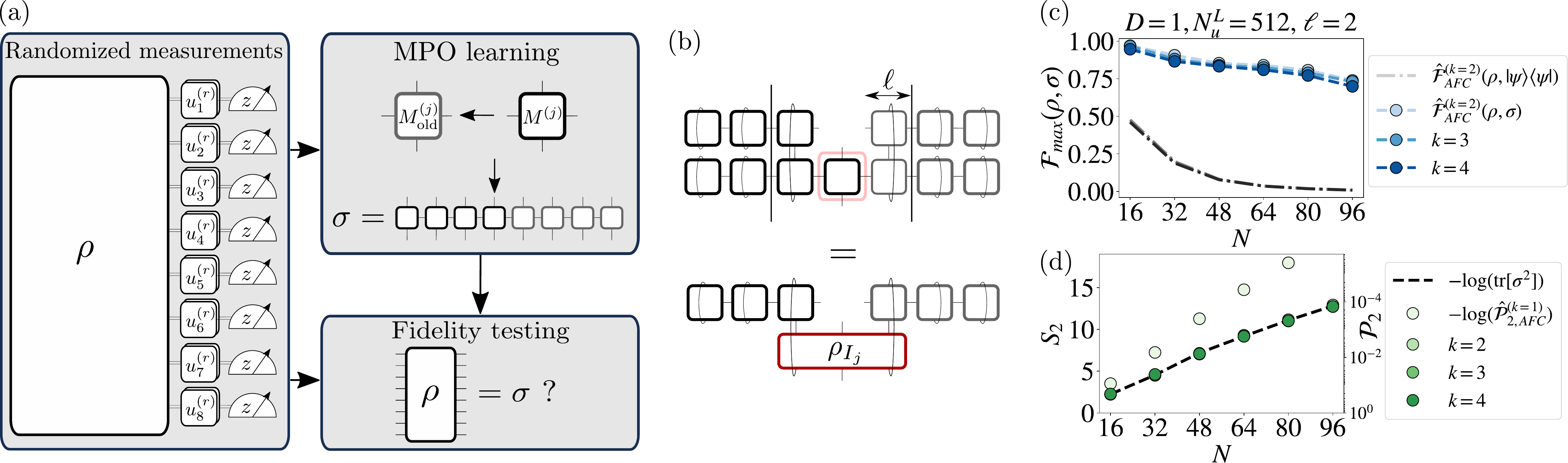}
    \caption{\textit{Learning protocol} 
    $-$ 
    (a) We present a protocol to learn the MPO representation $\sigma$ of an experimental quantum state $\rho$ from two randomized measurement datasets. We use the first dataset to optimize $\sigma$, and the second one to benchmark the output $\sigma$ by estimating the fidelity $\mathcal{F}_{\text{max}}(\rho,\sigma)$.
    (b) We optimize each tensor $M^{(j)}$ approximately by solving the linear system Eq.~\eqref{eq:localupdates}, represented diagrammatically.
    Note that it requires experimental data depending only on the reduced density matrix $\rho_{I_j}$, where $I_j = [j-\ell, j+\ell]$.
    (c) We experimentally demonstrate the protocol by learning the state of the first $N$ qubits out of $N_{\text{tot}} = 96$ qubits on a superconducting quantum processor, where $\rho$ is obtained by implementing Eq.~\eqref{eq:kicked_Ising} at depth $D = 1$.
    We set $(\ell, \chi^{\prime}) = (2, 4)$ in the learning algorithm, and use a total of $N_u \times N_M = 2048 \times 1024$ measurements. 
    We compare the AFC fidelity [Eq.~\eqref{eq:AFC_fidelity}] of the experimental state $\rho$ with the learned MPO $\sigma$ (marked lines) and with the target state $\ket{\psi}$ (dotted lines).
    (d) We estimate the Rényi entropy $S_2$ using both our protocol (dashed line) and the testing set using AFC (markers).
    }
\label{fig:1}
\end{figure*}

Probing the quantum state of an $N-$qubit system is a crucial yet non-trivial step in the successful implementation of many quantum simulation and quantum computation protocols~\cite{Blatt2012, Gross2017, Preskill2018, Schafer2020, Kjaergaard2020, Altman2021, Morgado2021, Monroe2021, Alexeev2021, Pelucchi2022, Burkard2023}.
In this context, randomized measurements provide an effective strategy to probe several physical properties of experimental quantum systems~\cite{Elben_2019_statistical, Elben_2022_review, Cieliski_2024_review}, such as entanglement~\cite{Brydges_2019, Elben2020ptmoments, Neven_2021_SRPTmoments, rath2021qfi, rath2023opent, carrasco2024r2, vermersch2024AFC} and fidelities~\cite{da_Silva_2011,Flammia_2011,Elben2020crossplatform, Zhu_2022_crossplatform, huang2024certifying}.
In particular, the framework of classical shadows~\cite{Huang_2020_shadows} has become a routine tool to post-process randomized measurement datasets.
This is due to 
($i$) provable performance guarantees for statistical errors, that allow us to reach considerable system sizes~\cite{da_Silva_2011, Flammia_2011, Huang_2020_shadows, vermersch2024AFC},
($ii$) minimal hardware requirements (applying one layer of single-qubit unitaries before measurements suffices)~\cite{Elben_2022_review}, and
($iii$) robustness to measurement errors~\cite{Koh2022robust, chen2021robust, vitale2024qfi}.
While classical shadows are state-agnostic, typical quantum states generated in noisy quantum devices admit simple descriptions.
For instance, matrix-product operators (MPOs) can accurately describe output states of one-dimensional noisy quantum circuits~\cite{Noh2020noisy, Cheng2021noisy}, as well as one-dimensional thermal states relevant to quantum simulation~\cite{Verstraete_2004_MPDO,kuwahara2021gibbs}, with only $O(N)$ parameters. 
However, it remains unclear how to efficiently reconstruct such quantum states from classical shadows at large system sizes.

In this Letter, we introduce a protocol 
to learn the MPO representation of quantum states in large-scale experiments, that inherits the key properties ($i,ii,iii$) of classical shadows. 
This allows us to probe global properties of quantum states with extensive entropies in experimental settings.
We show this by learning many-body states generated by a superconducting quantum processor up to $N = 96$ qubits, whereas randomized measurements have been previously implemented only up to $N = 13$~\cite{vitale2024qfi}.
The MPO representation provides direct access to several physical properties, without the need of tailored estimation formulas, nor to re-process the dataset as in previous approaches.
Leveraging this full mixed-state description, we also investigate and quantify the effect of experimental noise and demonstrate large-scale error mitigation.

The MPO representation of a quantum state $\sigma$ is specified by a set of $N$ tensors $M^{(j)}$ as
\begin{equation}
\label{eq:def_sigma}
    \sigma =\!\sum_{\{s_j\}, \{s'_j\}}\! M^{(1)}_{s_1, s_1'} M^{(2)}_{s_2, s_2'} ...~M^{(N)}_{s_N, s_N'} \ket{\{s_j\}}\bra{\{s_j'\}},
\end{equation}
where $\ket{\{s_j\}}=\otimes_{j=1}^N|s_j\rangle$, with $s_j = 0,1$, the computational basis corresponding to the Hilbert space of $N$ qubits, while $M^{(j)}_{s_j, s_j'}$ are $\chi_j \times \chi_{j+1}$ matrices. 
The bond dimension $\chi = \text{max}(\chi_j)$ controls the expressivity of this ansatz~\cite{Guth_Jarkovsk__2020_MPDO}: 
while any quantum state can be described by a MPO with $\chi = 2^N$, 
typical noisy~\cite{Noh2020noisy, Cheng2021noisy,Verstraete_2004_MPDO} or thermal~\cite{kuwahara2021gibbs} quantum states in one dimension only require $\chi = O(1)$.
Other approaches to quantum state tomography based on MPO have been proposed in recent years.
They consist of either 
numerical methods without analytical performance guarantees~\cite{Baumgratz2013maxlikelihood,lidiak2022TTcross, Torlai2023PastaQ, Guo_2024LPDO, jameson2024SIC}, or 
provably efficient proposals [($i$)]~\cite{baumgratz2013MPO, holzapfel2018petz, Qin_2024_MPDOtomo, qin2024designs, fanizza2024learningfcs,qin2025PCS}, but incompatible with local randomized measurements [($ii$)] or the robust classical shadow framework [($iii$)].
Similar considerations apply to protocols for the tomography of Gibbs states~\cite{Swingle_2014,Anshu_2021,Kokail_2021,Joshi_2023, Rouz__2024, Haah_2024, Bakshi_2024, capel2024conditionalindependence}, another candidate ansatz to learn mixed states in experiments~\cite{anshu2024survey}.
Here we obtain the properties ($i$, $ii$, $iii$) by combining bounds on statistical errors from the classical shadow framework, with the analytical properties and computational efficiency of MPOs.

Our protocol is illustrated in Fig.~\ref{fig:1}: we perform randomized measurements on the state $\rho$ in $N_u = N_u^L + N_u^T$ local random bases with $N_M$ shots per basis, and assign $N_u^L$ bases to the learning set, and $N_u^T$ bases to the testing set.
The learning set serves as input for a tensor network learning algorithm where each tensor of a MPO $\sigma$ is optimized sequentially, similarly to density-matrix renormalization group (DMRG) and related approaches~\cite{Schollw_ck_2011_dmrg, stoudenmire2016learning, han2018learning, ayral2023dmrg}.
Then, we use the testing set to benchmark the result by estimating the max fidelity~\cite{Liang_2019, Elben2020crossplatform} 
\begin{equation}
\label{eq:def_Fmax}
    \mathcal{F}_{\text{max}}(\rho, \sigma) = \frac{\text{tr}[\rho\sigma]}{\text{max}(\text{tr}[\rho^2], \text{tr}[\sigma^2])}.
\end{equation}
Importantly, we prove that the optimization of the tensors $M^{(j)}$ and the estimation of $\mathcal{F}_{\text{max}}(\rho, \sigma)$ can be performed with $O(\text{poly}(N))$ measurements under mild technical assumptions.
In the remainder of the manuscript we illustrate our protocol and its experimental implementation, together with a demonstration and discussions of possible applications.

\prlsection{Learning algorithm}
In order to illustrate and justify all the steps of our algorithm, we make some assumptions on the target state $\rho$ to be learned: 
($a$) $\rho$ is a MPO with a bond dimension $\chi$, and 
($b$) both the density matrices $\rho$ and $\rho^\prime\propto \rho^2$ have finite correlation lengths. 
The latter assumption allows for an efficient estimation of fidelities, as we comment later, and implies the following approximate factorization conditions (AFC) for the purity~\cite{vermersch2024AFC, capel2024conditionalindependence}
\begin{equation}
\label{eq:def_AFCpurity}
    \Bigg|\text{tr}[\rho^2_{ABC}]^{-1}\frac{\text{tr}[\rho^2_{AB}]\text{tr}[\rho^2_{BC}]}{\text{tr}[\rho^2_{B}]} - 1 \Bigg| \leq \alpha e^{-|B|/\xi_{\rho}^{(2)}},
\end{equation}
where $\alpha>0$ is a constant, $\xi_{\rho}^{(2)}>0$ is the correlation length of $\rho^\prime\propto \rho^2$, and $\rho_{I} = \text{tr}_{\bar{I}}[\rho]$, where $A$, $B$, and $C$ form a tripartition for a connected subset of qubits. 
The correlation length $\xi_{\rho}^{(2)}$ is also referred to as the Rényi-Markov length of $\rho$~\cite{sang2024markov, zhang2025renyimarkov}.
Note that $\xi_{\rho}^{(2)}$ is not related to the injectivity length of matrix-product states (MPSs)~\cite{Cirac_2021}, which is the relevant length scale in MPS tomography~\cite{Cramer2010MPStomo, Lanyon_2017}.
Later, we will comment on the applicability of our method when these assumptions are only approximately met. 

The learning algorithm starts from an initial MPO ansatz $\sigma$ with a given bond dimension $\chi'$.
Then, we update sequentially the tensors $M^{(j)}$ in Eq.~\eqref{eq:def_sigma} by maximizing the geometric-mean (GM) fidelity $\mathcal{F}_{\text{GM}}(\rho, \sigma) = \frac{\text{tr}[\rho\sigma]}{\sqrt{\text{tr}[\rho^2]\text{tr}[\sigma^2]}}$~\cite{Liang_2019, Elben2020crossplatform}.
Unlike the max fidelity, the GM fidelity is differentiable, but at the same time it is insensitive to certain types of decoherence~\cite{Elben2020crossplatform, SM}.
The key result of this work is that, if $\rho$ is a translation-invariant MPO satisfying the assumptions above, and given a parameter $\ell>O(\xi_{\rho}^{(2)})$, the tensor $M^{(j)}$ that maximizes the GM fidelity approximately satisfies~\cite{SM, koor2023wirtinger}
\begin{equation}
\label{eq:localupdates}
    \text{tr}_{I_j\setminus \{j\}}[\sigma_{I_j}\partial_{M^{(j)}}\sigma_{I_j}] = \text{tr}_{I_j\setminus  \{j\}}[\rho_{I_j}\partial_{M^{(j)}}\sigma_{I_j}],
\end{equation}
where the trace is taken over all the qubits in $I_j$ but $j$, and $I_j = [j-\ell, j+\ell]$, see Fig.~\ref{fig:1}(b) for a diagrammatic representation.
More specifically, when the algorithm is close to convergence, we can prove that $||M^{(j)} - M_0^{(j)}||_2 = O(\text{tr}[\rho_{I_j}\sigma_{I_j}]\chi e^{-\ell/2\xi_{\rho\sigma}})$, where $M_0^{(j)}$ is the tensor solving Eq.~\eqref{eq:localupdates}, and $\xi_{\rho\sigma}$ is the correlation length of the operator $\rho\sigma$.
This approximate update is convenient, since the right-hand side of Eq.~\eqref{eq:localupdates} can be estimated efficiently via classical shadows, as we show later.
For each tensor, Eq.~\eqref{eq:localupdates} can be solved numerically as a linear system in $M^{(j)}$ if $\chi^\prime\leq 4^{\ell}$~\cite{SM}.
The algorithm then proceeds by sweeping through the tensors $M^{(j)}$ from $j=1$ to $j={N}$ and back, performing a total of $N_S$ sweeps. 
After each sweep, we estimate $\mathcal{F}_{\text{max}}(\rho, \sigma)$, which is the control parameter of the algorithm.
Importantly, it is not necessary to know the values of $\xi_{\rho}^{(2)}$ and $\chi$ exactly: in order for the algorithm to accurately reconstruct $\rho$, it is only necessary that $\ell>O(\xi_{\rho}^{(2)})$ and $\chi^\prime\geq\chi$ (however, similarly to standard MPS tomography~\cite{Cramer2010MPStomo, Lanyon_2017, kurmapu2023mpslearning}, the number of measurements and post-processing operations increase with them). 
When $\rho$ is only approximately described by a MPO with bond dimension $\chi$, we could use this approach until we reach the desired accuracy, that will be constrained by truncation errors.

\prlsection{Randomized measurement protocol}
At this stage, we are left with the task of estimating ($a$) the right-hand side of Eq.~\eqref{eq:localupdates} and ($b$) $\mathcal{F}_{\text{max}}(\rho, \sigma)$ using respectively the learning and the testing set of randomized measurements obtained from $\rho$.
A key feature of randomized measurements is that we can re-evaluate both these functions for any $\sigma$ using the same dataset.
Using the classical shadow formalism~\cite{Huang_2020_shadows}, we can estimate $\text{tr}_{I_j\setminus  \{j\}}[\rho_{I_j}\partial_{M^{(j)}}\sigma_{I_j}]$ as $\frac{1}{N_u^L N_M}\sum_{r=1}^{N_u^L N_M}\text{tr}_{I_j\setminus  \{j\}}[\hat{\rho}_{I_j}^{(r)}\partial_{M^{(j)}}\sigma_{I_j}]$, where each
\begin{equation}
\label{eq:def_classicalshadows}
    \hat{\rho}^{(r)}_{I_j} = \bigotimes_{j\in I_j}\left( 3u^{(r)\dagger}_j\ket{s^{(r)}_j}\bra{s^{(r)}_j}u_j^{(r)} - \mathbb{I}_j\right)
\end{equation}
is a classical shadow of $\rho_{I_j}$.
Here, $\{s_j^{(r)} = 0, 1\}$ are the measurement outcomes in the computational basis, $u_j^{(r)}$ are the random unitary transformations applied to each qubit $j$, and $\mathbb{I}_j$ is the identity. 
Crucially, this estimation can be done in norm$-2$ error $\varepsilon$ with $N_u^L = O(2^{2\ell}{\chi'}\varepsilon^{-2})$ measurement bases~\cite{SM, AltomareCampiti+1994}.
This implies that statistical errors in learning each tensor $M^{(j)}$ can be made arbitrarily small with a number of measurements independent from the total number of qubits $N$.

We now propose a scheme to estimate $\mathcal{F}_{\text{max}}(\rho, \sigma)$ [Eq.~\eqref{eq:def_Fmax}] with a controlled measurement budget.
While its estimation generally requires $N_u^TN_M = O(4^{N})$ randomized measurements~\cite{Elben2020crossplatform, Zhu_2022_crossplatform, rath2021qfi}, it is possible to drastically reduce this overhead under the previous assumption that the correlation lengths $\xi_{\rho}^{(2)}$ and $\xi_{\rho\sigma}$ are finite.
In this case, we prove that~\cite{SM}
\begin{align}
\label{eq:def_AFCoverlap}
 \text{tr}[\rho\sigma]\! &\simeq\!   \mathcal{O}_{\text{AFC}}^{(k)}(\rho, \sigma)\!=\! \frac{\prod_{j=1}^{\lfloor N/k\rfloor-1}\!\text{tr}[\rho_{A_{j}A_{j+1}}\sigma_{A_{j}A_{j+1}}]}{\prod_{j=2}^{\lfloor N/k\rfloor-1}\!\text{tr}[\rho_{A_{j}}\sigma_{A_{j}}]},
\end{align}
where $A_j$ are neighboring regions with $|A_j|=k>O( \xi_{\rho\sigma})$.
The same assumptions yield a similar result for the purity, \textit{i.e.} $\text{tr}[\rho^2]\simeq \mathcal{P}^{(k)}_{{2, \rm AFC}}(\rho) = \mathcal{O}_{\text{AFC}}^{(k)}(\rho, \rho)$~\cite{vermersch2024AFC}.
Defining now the AFC max fidelity
\begin{equation}
\label{eq:AFC_fidelity}
    \mathcal{F}_{\text{AFC}}^{(k)}(\rho,\sigma) = \frac{\mathcal{O}_{\text{AFC}}^{(k)}(\rho, \sigma)}{\text{max}(\mathcal{P}_{2,\text{AFC}}^{(k)}(\rho), \mathcal{P}_{2,\text{AFC}}^{(k)}(\sigma))},
\end{equation}
Eq.~\eqref{eq:def_AFCoverlap} yields $\mathcal{F}_{\text{max}}(\rho,\sigma)\simeq\mathcal{F}_{\text{AFC}}^{(k)}(\rho,\sigma)$. 
This approximation finally allows us to estimate $\mathcal{F}_{\text{max}}(\rho,\sigma)$ efficiently. 
Indeed, $N_u^T N_M = O(4^{2k}N^3)$ measurements are sufficient to estimate $\mathcal{F}_{\text{AFC}}^{(k)}(\rho,\sigma)$ with finite statistical errors, while the systematic error $|\mathcal{F}_{\text{max}}(\rho,\sigma)- \mathcal{F}_{\text{AFC}}^{(k)}(\rho,\sigma)|$ decreases exponentially with $k$~\cite{SM, vermersch2024AFC}.
Now, we can test quantitatively the output $\sigma$ of the learning algorithm by estimating $\mathcal{F}_{\text{AFC}}^{(k)}(\rho, \sigma)$ for various $k\geq \ell$.
Note that this method relies on the assumption that $\rho$ satisfies AFC for some finite $k$, and, in contrast to MPS verification~\cite{Cramer2010MPStomo, Lanyon_2017}, the efficient verification of AFC is an open problem.

\begin{figure}[t]
    \includegraphics[width=0.48\textwidth]{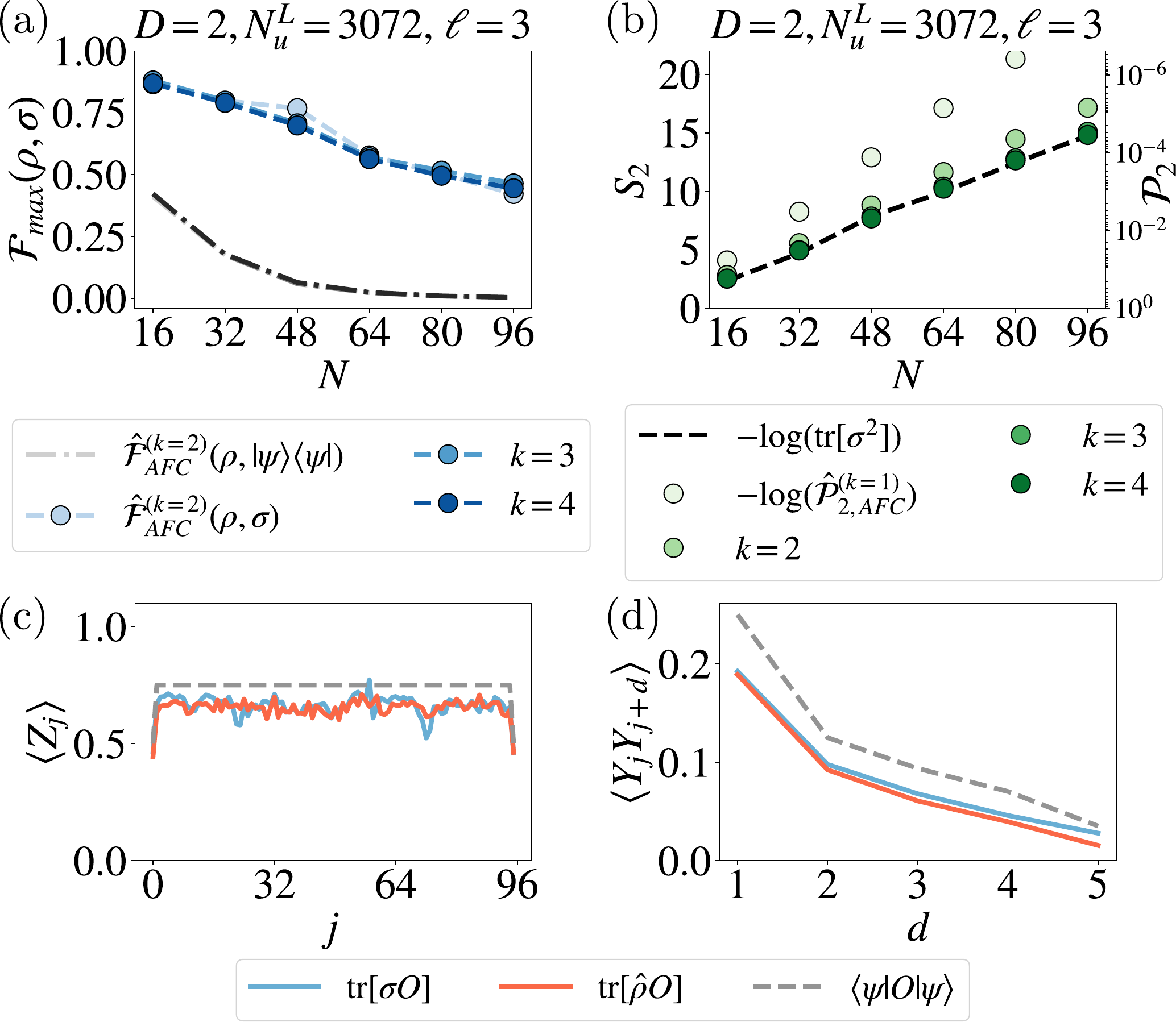}
    \caption{\textit{Experimental results} 
    $-$ 
    Results for a kicked Ising model with depth $D=2$, $N_u^L + N_u^T = 4096$ and $N_M = 1024$, and $(\ell, \chi^\prime) = (3, 8)$.
    (a) $\mathcal{F}_{\text{max}}$ as a function of $N$, with respect to the experimental quantum state (markers) and the ideal target state (dotted line).
    (b) Global purity as a function of $N$, obtained from the learned MPO $\sigma$ (dashed line) and from the testing set using Eq.~\eqref{eq:def_AFCoverlap} (markers).
    (c, d) Expectation values of the magnetization and $2-$body correlation functions (averaged over $j \in [3, 88]$), comparing values obtained from the learning protocol (blue lines), classical shadows from the testing set (red lines), and the pure state $\ket{\psi}$ (dashed lines). 
    }
\label{fig:2}
\end{figure}
\prlsection{Experimental results}
We now demonstrate the practical feasibility of our protocol by learning entangled quantum states from a partition of $N_{\text{tot}} = 96$ qubits of the IBM Brisbane superconducting quantum processing unit.
We prepare entangled states by running the kicked Ising model quantum circuit at low depth $D$, defined as~\cite{Kim2023}
\begin{equation}
\label{eq:kicked_Ising}
    \ket{\psi} = \left(\prod_{j=1}^{N_{\text{tot}}-1}e^{i\frac{\pi}{4} Z_jZ_{j+1}} \prod_{j=1}^{N_{\text{tot}}}e^{-i\frac{\pi}{8} X_j}\right)^D \ket{0}^{\otimes N},
\end{equation}
where $X_j, Y_j, Z_j$ are Pauli operators.
The preparation and the detection of the target state $\ket{\psi}$ are however affected by experimental noise, resulting in the observation of a mixed state $\rho$ which we analyze with our protocol.

For $D = 1$ and $D=2$, we perform local randomized measurements with $N_M = 1024$ shots per basis and $N_u = N_u^L + N_u^T = 2048\cdot D$ bases, and post-process the experimental data with the previously introduced learning algorithm.
As a slight modification, we perform two-site updates (instead of one-site updates) of the tensors $M^{(j)}$, as this allows us to dynamically increase the bond dimension up to a value $\chi^\prime$~\cite{SM}. 
In addition, we implement common randomized measurements in the estimation of Eq.~\eqref{eq:localupdates} to reduce statistical errors~\cite{vermersch2024common} (although this step does not drastically change the results~\cite{SM}).
We always initialize the algorithm in the ansatz $\sigma=(\mathbb{I}/2)^{\otimes N}$ and perform $N_S=20$ sweeps.

We report the results in Figs.~\ref{fig:1} and \ref{fig:2}.
First, we benchmark the results of the algorithm by comparing the fidelities $\mathcal{F}_{\text{max}}(\rho, \sigma)$ and $\bra{\psi}\rho\ket{\psi}$.
We observe that $\mathcal{F}_{\text{max}}(\rho, \sigma) \gg \bra{\psi}\rho\ket{\psi}$, meaning that the learned MPO $\sigma$ correctly captures the effect of experimental noise.
For $N = 96$ and $D = 1, 2$ we obtain fidelities of $\mathcal{F}_{\text{max}}(\rho,\sigma)\sim 75\%$ and $\sim 50\%$ respectively, mainly due to statistical errors (as we similarly observe in numerical examples~\cite{SM}).
In a product state, the observed values would correspond to a fidelity per qubit $(\mathcal{F}_{\text{max}}(\rho,\sigma))^{1/N}>99\%$.
To get a more meaningful insight, we study the magnetization $\langle Z_j \rangle$ and the two-body correlation functions $\langle Y_j Y_{j+d} \rangle$, see Fig.~\ref{fig:2}(c,d). 
We compute them for the states $\rho$, $\sigma$ and $\ket{\psi}$, using standard shadow estimation~\cite{Huang_2020_shadows} from the testing set for the former.
We observe that the expectation values obtained from $\rho$ and $\sigma$ are in good agreement, and both highlight non-ideal features of the experimental quantum state.

These observation corroborate that $\rho$ is well-described by a mixed state.
To quantify this feature, we now analyze the second Rényi entropy $S_2(\rho) = -\text{log}(\text{tr}[\rho^2])$~\footnote{We consider the logarithm to be in base 2, such that $S_2\left( (\mathbb{I}/2)^{\otimes N}\right) = N$.} of the full state, which can also be estimated using randomized measurements together with AFC~\cite{vermersch2024AFC}. 
In Figs.~\ref{fig:1}(d),\ref{fig:2}(b) we compare its AFC estimate $-\text{log}(\mathcal{P}^{(k)}_{2,\text{AFC}})$ from the testing set with $S_2(\sigma)$, observing that they are in perfect agreement at large $k$.
While previous experimental approaches were limited to $N=12,13$ qubits~\cite{vitale2024qfi, andersen2025thermalization}, here we measure large-scale entropies to a high degree of accuracy, that correspond to exponentially small purities.
By doing so we observe that
the entropy of these states obeys a weak-volume law $S_2\simeq \alpha N$, where the coefficient $\alpha\simeq 0.14-0.15$, compared to $\alpha = 1$ for a fully mixed state, 
and the modest increase of the global entropy from $D=1$ to $D=2$ suggests that readout errors are dominant in the low-depth regime.
Overall, these results demonstrate experimental tomography of $N = 96$ qubits entangled quantum states, going beyond the state-of-the-art of $N = 20$ qubits obtained for MPS~\cite{kurmapu2023mpslearning} and Gibbs states~\cite{Joshi_2023}.

\begin{figure}[t]
    \includegraphics[width=0.48\textwidth]{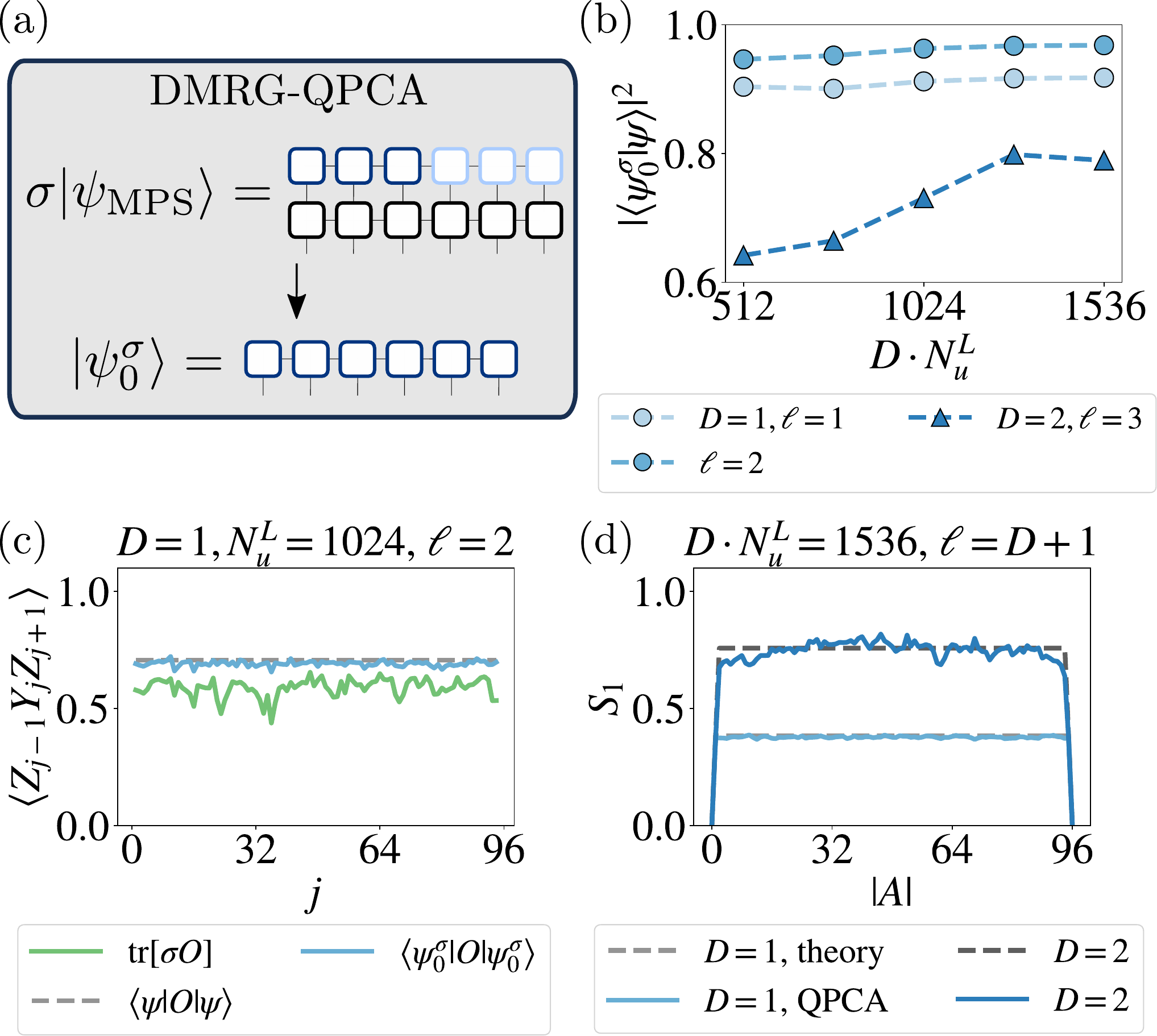}
    \caption{\textit{Experimental quantum principal component analysis} 
    $-$ 
    (a) Running the DMRG algorithm on $H = -\sigma$ allows us to approximately find the principal component $\ket{\psi_0^{\sigma}}$, which approximates the principal component of the experimental state $\rho$.
    (b) Fidelity between $\ket{\psi_0^{\sigma}}$ and the target state $\ket{\psi}$ for the experiment at depth $D = 1$, as a function of $\ell$ and $N_u$.
    (c) Expectation values of a local observable for depth $D = 1$, compared between the learned MPO $\sigma$ and the error-mitigated state to the theory prediction.
    (d) Bipartite Von Neumann entropy as a function of the cut for $D = 1, 2$, compared between $\ket{\psi_0^{\sigma}}$ and $\ket{\psi}$. 
    }
\label{fig:3}
\end{figure}

\prlsection{Application: quantum error mitigation}
We now leverage the access to a MPO description of $\rho$ to perform quantum error mitigation.
Quantum error mitigation consists in partially removing the effect of errors in the estimation of properties from a quantum experiment via classical post-processing~\cite{Cai_2023_mitigation}.
Conventional approaches are usually based on assuming specific noise models~\cite{van_den_Berg_2023_PEC}, or require additional experiments~\cite{giurgica-tiron2020ZNE,guo2022errormitigation,filippov2023TEM,Mangini_2024_TEM}.
Our approach consists instead in quantum principal component analysis (QPCA)~\cite{Lloyd_2014_QPCA, Huang_2022_learning}, where we mitigate experimental noise by reconstructing the pure state $\ket{\psi_0^{\rho}}$ with the largest eigenvalue in the spectral decomposition $\rho = \sum_a \Lambda_a \ket{\psi_a^{\rho}}\bra{\psi_a^{\rho}}$, where $\Lambda_a>\Lambda_{a+1}$~\cite{Koczor_2021_QPCA}.
While this idea is powerful, it has been implemented only approximately by virtual distillation~\cite{cotler2019cooling,huggins2021virtualcooling,seif2023virtualdistillation, Hakoshima_2024_virtualdistillation}.

Here we perform QPCA by running the DMRG algorithm~\cite{Schollw_ck_2011_dmrg} on the MPO $H = -\sigma$, obtaining the state $\ket{\psi_0^{\sigma}}$ as a MPS.
Note that this approach does not require additional experiments nor data analysis.
In Fig.~\ref{fig:3} we illustrate the results of DMRG-QPCA on the experimental data using $N = N_{\text{tot}}$~\footnote{We initialize DMRG with $\ket{\psi_0}\propto \ket{\psi_{\text{RMPS}}} + \epsilon\ket{0}^{\otimes N}$ where $\ket{\psi_{\text{RMPS}}}$ is a random MPS with small bond dimension, and $\epsilon\gtrsim 10^{-2}$. For $\epsilon = 0$, we numerically observe that the DMRG algorithm does not update the random MPS.}.
We obtain in this way $\ket{\psi_0^{\sigma}}$ for $D = 1, 2$, and their corresponding eigenvalues $\Lambda_0 = 9.11\cdot10^{-3}, 3.13\cdot10^{-3}$.
Remarkably, we obtain high fidelities between the MPSs $\ket{\psi_0^\sigma}$ and the target states $\ket{\psi}$, even larger that $90\%$ for $D = 1$, corroborating the power of this error mitigation method.
This is also reflected by expectation values of local observables, that show a dramatic improvement from the MPO prediction of $\sigma$.
Finally, we leverage the MPS representation to observe the growth of the bipartite Von Neumann entropy $S_1 = \text{tr}_{A}[\rho'~\text{log}~\rho']$ with the circuit depth, where we take $\rho' = \text{tr}_{\bar{A}}[\ket{\psi_0^{\sigma}}\bra{\psi_0^{\sigma}}]$ and $A = [1, |A|]$, see Fig.~\ref{fig:3}(d).
These results experimentally demonstrate QPCA at large scales, and provide a strong case for the application of our learning protocol to quantum error mitigation.

\prlsection{Conclusions and outlook}
Our protocol provides a quantum-to-classical converter~\cite{huang2022provably}, that compresses $N_u N_M$ randomized measurement outcomes from an experimental quantum state $\rho$ in a single classical object, the MPO $\sigma$.
Having access to an MPO approximation of $\rho$ not only allows us to estimate several physical properties efficiently, \textit{i.e.}~without re-processing experimental data each time, but also to apply powerful error-mitigation based on tensor-network algorithms.

Our ability to combine quantum experiments with classical tensor-network methods opens up further possibilities.
For instance, MPO tomography can be associated with MPS preparation algorithms which take as input an MPO/MPS $\sigma$ represented classically and devise the quantum circuit that efficiently forms the corresponding state $\rho$ in an universal quantum computer~\cite{Malz_2024, Fomichev2024chemistry}. 
This gives us the possibility to interact multiple times with a quantum system, or even connect different experiments to perform quantum computing beyond the possibilities of each individual device. 

An interesting use-case consists in quantum circuit cutting:
by saving an intermediate state of a quantum algorithm as a MPO and performing QPCA on it, one could resume the quantum algorithm by reloading the (error-mitigated) state with a MPS preparation circuit.
Another concrete application consists in modular experiments where a fault-tolerant quantum computer is used to analyze the state of a noisy quantum simulator.
For instance, one could
prepare a quantum state of interest in a quantum simulation experiment, and perform MPO tomography; next, one could
use a quantum circuit to prepare its MPS purification on a fault-tolerant quantum computer, and finally apply a quantum algorithm to measure the von Neumann entropy~\cite{gur2021sublinear,luongo2024determinant,giovannetti2025determinant}.

\let\oldaddcontentsline\addcontentsline
\renewcommand{\addcontentsline}[3]{}
\section*{Acknowledgements}

We acknowledge insightful discussions with Antoine Browaeys, Mari Carmen Bañuls, Soonwon Choi, Thierry Lahaye, Daniel Stilck-França, Georgios Styliaris, and Xavier Waintal.
The experimental data have been collected using the Qiskit library~\cite{javadiabhari2024qiskit}, and post-processed using the RandomMeas~\cite{RandomMeas} and ITensor~\cite{fishman2022itensor} libraries.
The work of M.V. and B.V. was funded by the French National Research Agency via the JCJC project QRand (ANR-20-CE47-0005), and via the research programs Plan France 2030 EPIQ (ANR-22-PETQ-0007), QUBITAF (ANR-22-PETQ-0004) and HQI (ANR-22-PNCQ-0002). 
We acknowledge the use of IBM Quantum Credits for this work. The views expressed are those of the authors, and do not reflect the official policy or position of IBM or the IBM Quantum team.
M.L. acknowledges support by the Deutsche Forschungsgemeinschaft (DFG, German Research Foundation) under Germany’s Excellence Strategy – EXC-2111 – 390814868. 
The work of C.L. was funded by the French National Research Agency via the PRC project ESQuisses (ANR-20-CE47-0014-01).
J.I.C. acknowledges funding from the Federal Ministry of Education and Research Germany (BMBF) via the project FermiQP (13N15889). Work at MPQ is part of the Munich Quantum Valley, which is supported by the Bavarian state government with funds from the Hightech Agenda Bayern Plus.
P.Z. acknowledges support by the European Union’s Horizon Europe research and innovation programme under grant agreement No 101113690 (PASQANS2).
The work of L.P. was funded by the European Union (ERC, QUANTHEM, 101114881). Views and opinions expressed are however those of the author(s) only and do not necessarily reflect those of the European Union or the European Research Council Executive Agency. Neither the European Union nor the granting authority can be held responsible for them.
We acknowledge support by the Erwin Schrödinger International Institute for Mathematics and
Physics (ESI).

\bibliography{MPO}
\let\addcontentsline\oldaddcontentsline

\onecolumngrid
\newpage

\appendix
\setcounter{equation}{0}
\setcounter{figure}{0}
\setcounter{page}{1}
\renewcommand{\thefigure}{S\arabic{figure}}

\setcounter{secnumdepth}{2}

\begin{center}
    {\large \bf Supplemental Materials: Learning mixed quantum states in large-scale experiments}
\end{center}

\tableofcontents

\section{Learning algorithm}
\label{app:algorithm}

In this section we derive the optimization step of the learning algorithm described in the main text.
In particular, given an experimental state $\rho$ to be learned and a MPO ansatz $\sigma$, defined as
\begin{equation}
\label{eqapp:def_MPO}
    \sigma = \sum_{\{s_j\}, \{s'_j\}} M^{(1)}_{s_1, s_1'} M^{(2)}_{s_2, s_2'} ...~M^{(N)}_{s_N, s_N'} \ket{\{s_j\}}\bra{\{s_j'\}} \ ,
\end{equation}
we prove that the optimization step for the tensor $M^{(j)}$ proposed in the main text approximately maximizes the geometric-mean fidelity~\cite{Liang_2019,Elben2020crossplatform} 
\begin{equation}
\label{eqapp:GM_fidelity}
    \mathcal{F}_{\text{GM}}(\rho, \sigma) = \frac{\text{tr}[\rho\sigma]}{\sqrt{\text{tr}[\rho^2]\text{tr}[\sigma^2]}} \ ,
\end{equation}
that is differentiable and can be efficiently computed numerically.
This is proven in two steps.
First, we derive the gradient of $\mathcal{F}_{\text{GM}}(\rho, \sigma)$ with respect to a tensor $M^{(j)}$ of the MPO $\sigma$, and show that the optimization step can be carried out exactly.
This step is similar to the derivation of the optimization step in the density matrix renormalization group (DMRG) algorithm~\cite{Schollw_ck_2011_dmrg} and other sequential tensor network optimization schemes~\cite{stoudenmire2016learning,han2018learning,ayral2023dmrg}.
Then, we show how under the assumptions that ($a$) $\rho$ is a MPO with bond dimension $\chi$ and ($b$) the states $\rho$ and $\rho'\propto\rho^2$ have finite correlation lengths, the optimization can be carried out approximately.
This is due to an approximation of the two terms in the gradient, which we prove in detail in App.~\ref{app:proof_local_approx_update}.
Finally, we explain how to choose the parameters of the learning algorithm to guarantee that the optimization step leads to an accurate solution.

\subsection{Exact tensor optimization}

\begin{figure}[t]
    \centering
    \includegraphics[width=0.9\linewidth]{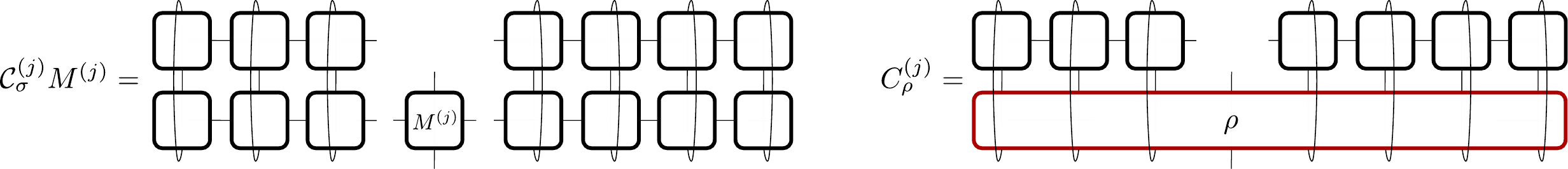}
    \caption{
    Diagrammatic representation of $\mathcal{C}_{\sigma}^{(j)}M^{(j)}$ and $C_{\rho}^{(j)}$ (defined in Eq.~\eqref{eqapp:def_C}) in terms of $\rho$ and the local tensor of $\sigma$. 
    }
    \label{fig:S1}
\end{figure}

We start by deriving the equation satisfied by an individual tensor $M^{(j)}$ of the MPO $\sigma$ that maximizes $\mathcal{F}_{\text{GM}}(\rho, \sigma)$.
As a first step, we compute the logarithmic gradient of the GM fidelity with respect to a local tensor 
\begin{equation}
\label{eqapp:gradient_def}
    G^{(j)} = \frac{\partial_{{M}^{(j)}}\mathcal{F}_{\text{GM}}(\rho, \sigma)}{\mathcal{F}_{\text{GM}}(\rho, \sigma)} = \frac{\text{tr}[\rho\partial_{{M}^{(j)}} \sigma]}{\text{tr}[\rho\sigma]} - \frac{\text{tr}[\sigma\partial_{{M}^{(j)}} \sigma]}{\text{tr}[\sigma^2]} \ ,
\end{equation}
where we have used Wirtinger derivative rules~\cite{koor2023wirtinger} to obtain that
\begin{equation}
    \partial_{M^{(j)}}\text{tr}[\rho\sigma] = \text{tr}[\rho\partial_{M^{(j)}}\sigma], \ \partial_{M^{(j)}}\text{tr}[\sigma^2] = 2\text{tr}[\sigma\partial_{M^{(j)}}\sigma]  \ .
\end{equation}
Assuming that the GM fidelity has only global maxima as a function of $M^{(j)}$, the optimal $M^{(j)}$ satisfies $G^{(j)}=0$.
We show now that it is possible to obtain the local tensor $M^{(j)}$ that satisfies this condition exactly by solving a linear system.
Let us define the terms
\begin{equation}
\label{eqapp:def_C}
    C^{(j)}_{\rho} = \text{tr}[\rho\partial_{{M}^{(j)}} \sigma], \ \mathcal{C}_{\sigma}^{(j)} M^{(j)} = \text{tr}[\sigma\partial_{{M}^{(j)}} \sigma] 
\end{equation}
such that both $C^{(j)}_{\rho}$ and $\mathcal{C}_{\sigma}^{(j)}$ do not depend on $M^{(j)}$ (see Fig.~\ref{fig:S1} for their diagrammatic notation).
In terms of these objects, the solution to $G^{(j)} = 0$ is given by
\begin{equation}
\label{eqapp:grad_sol}
    \mathcal{C}^{(j)}_{\sigma} M^{(j)} = \frac{\text{tr}[\sigma^2]}{\text{tr}[\rho\sigma]}C^{(j)}_{\rho} \ .
\end{equation}
Using that
\begin{equation}
    \text{tr}[\rho\sigma] = C_{\rho}^{(j)}\cdot M^{(j)} \ , \ \text{tr}[\sigma^2] = \mathcal{C}_{\sigma}^{(j)}M^{(j)}\cdot M^{(j)} 
\end{equation}
it is possible to see that Eq.~\eqref{eqapp:grad_sol} is solved by any tensor $M^{(j)}$ of the form
\begin{equation}
    M^{(j)} = \lambda \mathcal{C}_{\sigma}^{(j)-1}C_{\rho}^{(j)}
\end{equation}
for any coefficient $\lambda \neq 0$.
We initially choose $\lambda = 1$, which corresponds to imposing
\begin{equation}
\label{eqapp:condition_global}
    \text{tr}[\sigma\partial_{M^{(j)}}\sigma] = \text{tr}[\rho\partial_{M^{(j)}}\sigma] \ ,
\end{equation}
and then normalize the tensor $M^{(j)}$ such that $\text{tr}[\sigma]=1$.
We express the resulting update rule for the tensor $M^{(j)}$ in the following way
\begin{equation}
\label{eqapp:update_rule_global}
    M^{(j)}\leftarrow \mathcal{C}^{(j)-1}_{\sigma} C_{\rho}^{(j)}, ~ M^{(j)} \leftarrow \frac{1}{\text{tr}[\sigma]}M^{(j)} \ .
\end{equation}
where the first update rule consists in solving a linear system of ${4\chi'}^2$ equations, where $\chi'$ is the bond dimension of the MPO $\sigma$, and the second one normalizes $\sigma$.
This update rule also imposes the Hermiticity of $\sigma$, but not its positivity.
Finally, note that such an optimization scheme would require us to estimate the right-hand side of Eq.~\eqref{eqapp:condition_global} from experimental data, which requires $O(2^N)$ measurements using classical shadow framework~\cite{Huang_2020_shadows}.

\subsection{Approximate tensor optimization}

\begin{figure}[t]
    \centering
    \includegraphics[width=0.95\linewidth]{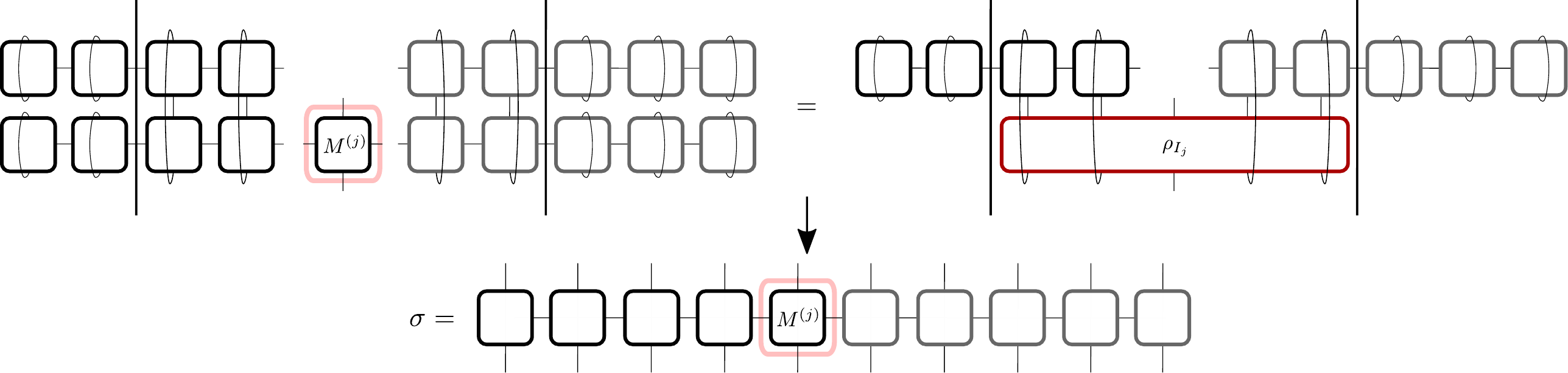}
    \caption{
    Graphical representation of the condition Eq.~\eqref{eqapp:approx_update_mt} for the approximate update rule for $N = 10$, $\ell = 2$, and $j = 5$. We are considering a right sweep, where the black tensors of $\sigma$ have been already updated, while the gray tensors not yet. 
    }
    \label{fig:S2}
\end{figure}

We now show that it is possible to approximately optimize the tensor $M^{(j)}$ as
\begin{equation}
\label{eqapp:approx_update_mt}
    \text{tr}[\sigma_{I_j}\partial_{M^{(j)}}\sigma_{I_j}] = \text{tr}[\rho_{I_j}\partial_{M^{(j)}}\sigma_{I_j}]
\end{equation}
up to a normalization factor.
This can be proven rigorously under the assumptions that
\begin{enumerate}
    \item $\rho$ is a translation-invariant matrix-product density operator (MPDO, i.e. a quantum state described by a MPO) with finite bond dimension $\chi$ and Rényi-Markov length $\xi_{\rho}^{(2)}$;
    \item the algorithm is close to convergence, so that we can assume that $\sigma$ is a translation-invariant MPDO with bond dimension $\chi$ as well.
\end{enumerate}
Note that the condition on the Rényi-Markov length is satisfied by typical MPDOs~\cite{Cirac_2021, vermersch2024AFC}.
Our starting point is the following lemma: 
given $\rho$ and $\sigma$ two translation-invariant MPDOs with bond dimension $\chi$, and given that the correlation length $\xi_{\rho\sigma}$ of the operator $\rho\sigma$ is finite, under mild technical assumptions the following approximation is valid (see App.~\ref{app:proof_local_approx_update} for a rigorous proof)
\begin{equation}
\label{eqapp:local_app_C}
    C_{\rho}^{(j)} = a_{\rho} (C_{\rho_{I_j}}^{(j)} + \varepsilon_{\rho}) \ , 
\end{equation}
with
\begin{equation}
    a_{\rho} = \frac{\text{tr}[\rho\tilde{\sigma}]}{\text{tr}[\rho_{I_j}\tilde{\sigma}_{I_j}]} \ ,
\end{equation}
where $I_j = [j-\ell, j+\ell]$, and $\tilde{\sigma}$ is the MPO obtained by substituting the tensor $M^{(j)}$ in $\sigma$ with any tensor $\tilde{M}^{(j)}$ such that $\text{tr}[\rho\tilde{\sigma}]\neq 0$, and 
\begin{equation}
\label{eqapp:local_app_C_err}
    ||\varepsilon_{\rho}||^2_2 \leq \text{tr}[\rho_{I_j}\tilde{\sigma}_{I_j}]^2 K\chi^2 e^{-\ell / \xi_{\rho\sigma}} \ ,
\end{equation}
where $K=O(1)$ is a constant.
Analogously, if the Rényi-Markov length $\xi_{\sigma}^{(2)}$ is also finite, the same is valid upon the substitution $\rho\rightarrow\sigma$, yielding
\begin{equation}
\label{eqapp:local_app_Cs}
    \mathcal{C}_{\sigma}^{(j)}\tilde{M}^{(j)} = a_{\sigma} (\mathcal{C}_{\sigma_{I_j}}^{(j)}\tilde{M}^{(j)} + \varepsilon_{\sigma}) \ ,
\end{equation}
for any tensor $\tilde{M}^{(j)}$ such that $\text{tr}[\tilde{\sigma}^2]\neq 0$.

Let us now take the equation satisfied by the optimal tensor $M^{(j)}$ [Eq.~\eqref{eqapp:grad_sol}]
\begin{equation}
    \mathcal{C}_{\sigma}^{(j)}M^{(j)} = \frac{M^{(j)}\cdot \mathcal{C}_{\sigma}^{(j)}M^{(j)}}{M^{(j)}\cdot C_{\rho}^{(j)}} C_{\rho}^{(j)} \ ,
\end{equation}
it is now possible to expand it by applying the approximations Eqs.~\eqref{eqapp:local_app_C},\eqref{eqapp:local_app_Cs}, obtaining
\begin{equation}
\label{eqapp:grad_sol_expanded}
    \mathcal{C}_{\sigma_{I_j}}^{(j)}M^{(j)} + \varepsilon_{\sigma} = \frac{M^{(j)} \cdot \mathcal{C}_{\sigma_{I_j}}^{(j)} M^{(j)} + M^{(j)} \cdot\varepsilon_{\sigma}}{M^{(j)} \cdot C_{\rho_{I_j}}^{(j)} + M^{(j)} \cdot\varepsilon_{\rho}}\left(C_{\rho_{I_j}}^{(j)} + \varepsilon_{\rho}\right) \ .
\end{equation}
By taking $||\varepsilon_{\rho}||_2,||\varepsilon_{\sigma}||_2 \rightarrow 0$, we can expand perturbatively Eq.~\eqref{eqapp:grad_sol_expanded} in $\varepsilon_{\rho}, \varepsilon_{\sigma}$, obtaining at the leading order
\begin{equation}
    \mathcal{C}_{\sigma_{I_j}}^{(j)}M_0^{(j)}  = \frac{M_0^{(j)} \cdot \mathcal{C}_{\sigma_{I_j}}^{(j)}M_0^{(j)} }{M_0^{(j)} \cdot C_{\rho_{I_j}}^{(j)} }C_{\rho_{I_j}}^{(j)}  \ ,
\end{equation}
where $||M^{(j)}-M_0^{(j)}||_2 = O(\varepsilon) $ with $\varepsilon = \text{max}(||\varepsilon_{\rho} ||_2, ||\varepsilon_{\sigma} ||_2)$.
If we assume that, close to convergence, $\xi_{\rho\sigma},\xi_{\sigma}^{(2)}\simeq\xi_{\rho}^{(2)}$, then for $\ell>O(\xi_{\rho}^{(2)})$, the optimal tensor $M^{(j)}$ can be approximated by the tensor $M_0^{(j)}$.
By repeating the same steps as in the previous section, it is possible to express $M_0^{(j)}$ as
\begin{equation}
    M_0^{(j)} = \lambda \mathcal{C}_{\sigma_{I_j}}^{(j)-1}C_{\rho_{I_j}}^{(j)}
\end{equation}
where the constant $\lambda$ is fixed by the normalization of $\sigma$.
This implies that we can first get $M_0^{(j)}$ for $\lambda = 1$ by solving Eq.~\eqref{eqapp:approx_update_mt}, and then fix $\lambda = \frac{1}{\text{tr}[\sigma]}$.
Therefore, the learning algorithm with approximate update is controlled by two parameters $(\ell, \chi')$, where the former is related to the Rényi-Markov length $\xi_{\rho}^{(2)}$, while the latter to the amount of entanglement encoded in $\rho$~\cite{Guth_Jarkovsk__2020_MPDO,Noh2020noisy, rath2023opent}.

\subsection{Convergence guarantees and two-site updates}

We now discuss for what choice of $(\ell, \chi')$ the approximate optimization step described above is guaranteed to converge.
As described in the main text, each tensor $M^{(j)}$ is optimized individually by solving Eq.~\eqref{eqapp:approx_update_mt}, which amounts to a linear system in $M^{(j)}$.
This linear system has $4{\chi'}^2$ variables, and at most $4(4^{\ell})^2$ independent equations.
This can be seen by performing a sequential singular-value decomposition (SVD) of $\sigma_{I_j}$, which results in a MPO with maximal bond dimension $4^{\ell}$.
This implies that the linear system solver is guaranteed to converge to a unique solution only if
\begin{equation}
    \chi' \leq 4^{\ell} \ .
\end{equation}
When this is not the case, some parameters of the local tensor $M^{(j)}$ may not be updated in a way which minimizes the global fidelity, resulting in the propagation of errors.

Another possibility consists in performing two-site updates, where at each step we optimize a tensor $\theta^{(j, j+1)} = M^{(j)}M^{(j+1)}$.
In this way, we do not have to fix the bond dimension of $\sigma$ from the beginning, but we can increase it dynamically, similarly to what is done in DMRG~\cite{Schollw_ck_2011_dmrg}.
However, even if the bond dimension is initialized to $\chi' = 1$, it can grow until $\chi' = 4^{\ell + 1}$, as the left and right bond dimensions of $\theta^{(j, j+1)}$ can grow up to $4^{\ell}$.
This means that we need to introduce a cut-off to the bond dimension $\chi'_{\text{max}} \leq 4^{\ell}$ when performing the SVD of $\theta^{(j, j+1)} = M^{(j)}M^{(j+1)}$ after a two-site update.
Otherwise, certain local updates might involve solving an under-determined linear system, leading to a possible propagation of uncontrolled errors.

\section{Local approximation of gradient terms}
\label{app:proof_local_approx_update}

In this section we provide a rigorous proof for the lemma Eq.~\eqref{eqapp:local_app_C_err} in the case of translation invariant MPDOs, establishing the validity of the local approximate updates with sufficiently large $\ell$.
Before doing so, we define translation-invariant matrix-product density operators (TIMPDOs), and we relate their properties to their transfer operators.
In particular, we rigorously define the correlation length of a MPDO.
The same language will be used in Sec.~\ref{app:AFC_fidelities} to prove approximation factorization conditions for the fidelity.

\subsection{Translation-invariant matrix-product operators and transfer operators}

\begin{figure}
    \centering
    \includegraphics[width=0.37\linewidth]{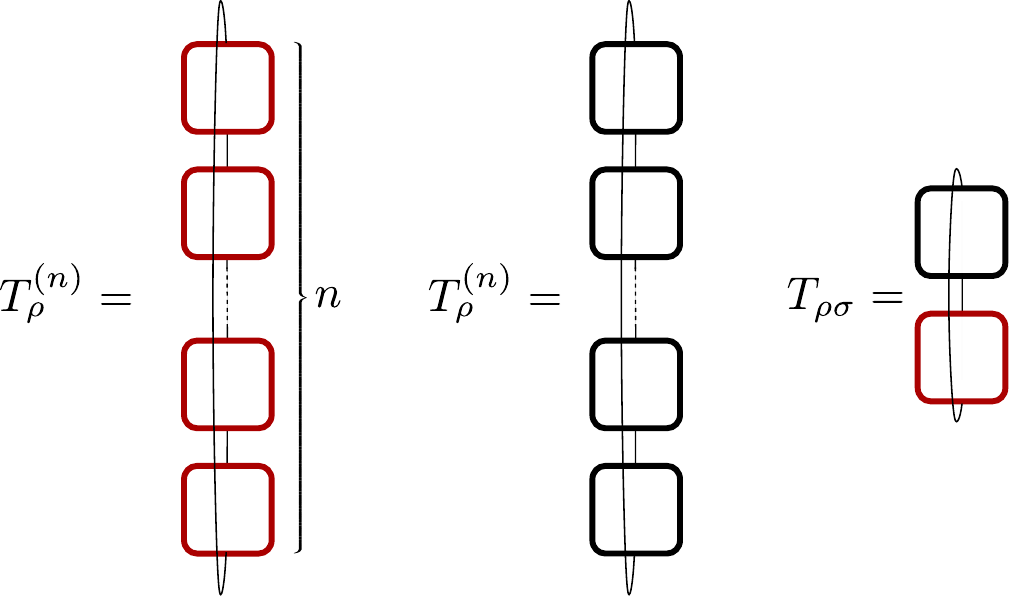}
    \caption{Graphical representation of the transfer operators $T_{\rho}^{(n)}$, $T_{\sigma}^{(n)}$ and $T_{\rho\sigma}$, defined in Eqs.~\eqref{eqapp:transfer_op_def1},\eqref{eqapp:transfer_op_def2}.}
    \label{fig:transfer_op}
\end{figure}

A quantum state $\rho$ is a translation-invariant MPDO (TIMPDO) if it can be written as in Eq.~\eqref{eqapp:def_MPO} and $M^{(j)} = M$, i.e. when all the tensors are equal.
To have a well-defined translation invariance, we consider MPDOs with periodic boundary conditions, i.e.
\begin{equation}
\label{eqapp:def_TIMPDO}
    \rho = \sum_{\{s_i\}, \{s'_i\}} \text{tr}[M'_{s_1, s_1'} M'_{s_2, s_2'} ...~ M'_{s_N, s_N'}] \ket{s_1, ...~, s_N}\bra{s'_1, ...~, s'_N} \ ,
\end{equation}
and similarly for $\sigma$
\begin{equation}
    \sigma = \sum_{\{s_j\}, \{s'_j\}} \text{tr}[M_{s_1, s_1'} M_{s_2, s_2'} ...~M_{s_N, s_N'}] \ket{s_1, ...~, s_N}\bra{s'_1, ...~, s'_N}  \ .
\end{equation}
For two TIMPDOs $\rho$ and $\sigma$ we can define the transfer operators (see Fig.~\ref{fig:transfer_op} for a graphical representation)
\begin{equation}
\label{eqapp:def_TM}
    T_{\rho}^{(n)} = \sum_{\{s_i^{(k)}\}} M'_{s_i^{(1)}, s_i^{(2)}}M'_{s_i^{(2)}, s_i^{(3)}} ... M'_{s_i^{(n)}, s_i^{(1)}} , \ T_{\sigma}^{(n)} = \sum_{\{s_i^{(k)}\}} M_{s_i^{(1)}, s_i^{(2)}}M_{s_i^{(2)}, s_i^{(3)}} ... M_{s_i^{(n)}, s_i^{(1)}} \ ,
\end{equation}
from which we can obtain the moments
\begin{equation}
\label{eqapp:transfer_op_def1}
    \text{tr}\left[\left(T^{(n)}_{\rho}\right)^N\right] = \text{tr}[\rho^n], ~ \text{tr}\left[\left(T^{(n)}_{\sigma}\right)^N\right] = \text{tr}[\sigma^n] \ .
\end{equation}
Additionally, we define the mixed transfer operator
\begin{equation}
\label{eqapp:transfer_op_def2}
    T_{\rho\sigma} = \sum_{s_i, s_i'} M^{(i)}_{s_i, s_i'} {M'}{}^{(i)}_{s_i', s_i} \ ,
\end{equation}
which analogously yields 
\begin{equation}
    \text{tr}[\rho\sigma] = \text{tr}\left[T_{\rho\sigma}^N\right] \ . 
\end{equation}
In what follows, we will focus on the spectral properties of the transfer operators, as obtained from the decompositions
\begin{equation}
    T_{\rho}^{(1)} = \sum_a \lambda_a \ket{l^{\rho}_a}\bra{r^{\rho}_a},\ T_{\sigma}^{(1)} = \sum_a \mu_a \ket{l^{\sigma}_a}\bra{r^{\sigma}_a},\ T_{\rho\sigma} = \sum_a \nu_a \ket{l^{\rho\sigma}_a}\bra{r^{\rho\sigma}_a} \ ,
\end{equation}
where the eigenvalues are ordered in absolute value (e.g. $|\lambda_i|\geq |\lambda_{i+1}|$), and we have the orthonormality conditions for left- and right-eigenvectors $\bra{l^O_a}r_b^O\rangle = \bra{r^O_a}l_b^O\rangle = \delta_{ab}$.
In case the spectrum of the transfer operator $T_{\rho}^{(1)}$ is gapped, i.e. $|\lambda_1/\lambda_0|<1$, the connected two-point correlation functions of $\rho$ will decay exponentially with the distance~\cite{Cirac_2021}, inducing the following definition of correlation length
\begin{equation}
    |\lambda_1/\lambda_0| = e^{-1/\xi_{\rho}^{(1)}} \ .
\end{equation}
Additionally, when a higher order transfer operator $T_{\rho}^{(n)}$ is gapped, it has been shown that the global entanglement properties (like the R\'enyi entropies) are approximately determined by subsystems of a finite length $O(\xi^{(n)}_{\rho})$, defined in an analogous way~\cite{vermersch2024AFC}.
This is also known as approximate conditional independence in terms of Rényi entropies~\cite{capel2024conditionalindependence}, and $\xi_{\rho}^{(n)}$ is called Rényi-Markov length~\cite{sang2024markov, zhang2025renyimarkov}.

In what follows, we assume the correlation lengths $\xi^{(1)}_{\rho}$, $\xi^{(2)}_{\rho}$ to be finite.
Additionally, we also assume the mixed transfer $T_{\rho\sigma}$ operator to be gapped, i.e. $\xi_{\rho\sigma} = \frac{1}{\text{log}|\nu_0/\nu_1|} = O(1)$.
While this last condition is much less explored, we believe it to be true for a generic pair of TIMPDOs with finite correlation lengths.
Practically, towards the convergence of the learning algorithm $\rho \sim \sigma$, we expect $\xi_{\rho\sigma}\sim \xi_{\rho}^{(2)} = O(1)$.
We make the further technical assumptions that the complex scalars 
\begin{equation}
    \Gamma^{bc}_a = \bra{r^{\rho\sigma}_a}l_b^{\rho} l^{\sigma}_c\rangle, \ \tilde{\Gamma}_{bc}^a = \bra{r_b^{\rho} r^{\sigma}_c} l^{\rho\sigma}_a\rangle 
\end{equation}
are both non-zero, i.e. $\Gamma_a^{bc}, \tilde{\Gamma}_{bc}^a \neq 0$.
We also believe this assumption to be true in the generic case~\cite{vermersch2024AFC}.

\subsection{Proof for local approximation of gradient terms}

In this section we use the transfer operator formalism to prove Eq.~\eqref{eqapp:local_app_C_err}.
Let us now take $\tilde{\sigma}$ to be the MPO obtained by substituting the tensor $M^{(j)}$ with an arbitrary tensor $\tilde{M}^{(j)}$ such that $\text{tr}[\rho\tilde{\sigma}]\neq 0$, where $\sigma$ is a TIMPDO with bond dimension $\chi$.
Furthermore, let us assume $\rho$ is a TIMPDO with bond dimension $\chi$ as well.
Then, under the assumptions that $\xi_{\rho\sigma} = O(1)$ and $\Gamma_a^{bc}, \tilde{\Gamma}_{bc}^a \neq 0$, we have
\begin{equation}
    \Bigg|\Bigg|\frac{C_{\rho_{I_j}}^{(j)}}{\text{tr}[\rho_{I_j}\tilde{\sigma}_{I_j}]}  - \frac{C_{\rho}^{(j)}}{\text{tr}[\rho\tilde{\sigma}]}\Bigg|\Bigg|_2^2 \leq K \chi^2 e^{-\ell / \xi_{\rho\sigma}} ~,
\end{equation}
where $C_{\rho}^{(j)} = \text{tr}[\rho\partial_{M^{(j)}}\sigma]$, $C_{\rho_{I_j}}^{(j)} = \text{tr}[\rho_{I_j}\partial_{M^{(j)}}\sigma_{I_j}]$ and $K=O(1)$ is a constant, and the norm$-2$ is the Hilbert-Schmidt norm
\begin{equation}
\label{eqapp:vector_norm}
    ||C_{\rho}^{(j)}||_2^2 = \sum_{a,a',s,s'} [\bar{C}_{\rho}^{(j)}]_{(a,a'),(s,{s'})} [C_{\rho}^{(j)}]_{(a,a'),(s,{s'})} \ ,
\end{equation}
where $a$, $a'$ ($s$, $s'$) are the two link (site) indices, and $\bar{C}_{\rho}^{(j)}$ is the complex conjugate of $C_{\rho}^{(j)}$. 
A proof similar to the following can be used to show that
\begin{equation}
    \Bigg|\Bigg|\frac{\mathcal{C}_{\sigma_{I_j}}^{(j)}\tilde{M}^{(j)}}{\text{tr}[\tilde{\sigma}_{I_j}^2]}  - \frac{\mathcal{C}_{\sigma}^{(j)}\tilde{M}^{(j)}}{\text{tr}[\tilde{\sigma}^2]}\Bigg|\Bigg|_2^2 \leq K' \chi^2 e^{-\ell / \xi_{\sigma}^{(2)}} \ ,
\end{equation}
where $\tilde{M}^{(j)}$ is any tensor such that $\text{tr}[\tilde{\sigma}^2]$, and $K'$ is a constant.

As a first step, let us expand the norm$-2$ distance as follows
\begin{equation}
\label{eqapp:norm2d_expanded}
    \Bigg|\Bigg|\frac{C_{\rho}^{(j)}}{\text{tr}[\rho\tilde{\sigma}]} - \frac{C_{\rho_{I_j}}^{(j)}}{\text{tr}[\rho_{I_j}\tilde{\sigma}_{I_j}]}\Bigg|\Bigg|_2^2 = \frac{\bar{C}_{\rho}^{(j)}\cdot C_{\rho}^{(j)}}{\text{tr}[\rho\tilde{\sigma}]^2} - \frac{2\text{Re}(\bar{C}_{\rho}^{(j)}\cdot C_{\rho_{I_j}}^{(j)})}{\text{tr}[\rho\tilde{\sigma}]\text{tr}[\rho_{I_j}\tilde{\sigma}_{I_j}]} + \frac{\bar{C}_{\rho_{I_j}}^{(j)}\cdot C_{\rho_{I_j}}^{(j)}}{\text{tr}[\rho_{I_j}\tilde{\sigma}_{I_j}]^2} \ .
\end{equation}
By expressing every factor in Eq.~\eqref{eqapp:norm2d_expanded} in terms of the spectral decompositions of the transfer operators, we obtain that each on of the three terms can be further decomposed in a term constant in $\ell$, and a term dependent on it.
As we now prove in two steps, the constant contribution is equal among the three terms and gets canceled out in the algebraic sum, while the other contribution admits an upper bound proportional to $O(\chi^2e^{-\ell/\xi_{\rho\sigma}})$.

\subsubsection{Spectral decompositions}

Using the spectral decompositions of the transfer operators, we obtain the following expressions for the overlaps
\begin{equation}
\label{eqapp:TO_overlaps}
\begin{split}
    \text{tr}[\rho\tilde{\sigma}] = \sum_{a=0}^{\chi^2-1} t_{aa}\nu_a^{N-1} ~,~
    \text{tr}[\rho_{I_j} \tilde{\sigma}_{I_j}] = \sum_{a,b=0}^{\chi^2-1}\sum_{c,d=0}^{\chi-1} t_{ab}\nu_a^{\ell}\nu_b^{\ell}\Gamma_b^{cd}(\lambda_c \mu_d)^{N-(2\ell+1)}\tilde{\Gamma}^a_{cd} \ , 
\end{split}
\end{equation}
where we have used $|I_j| = 2\ell+1$, and defined
\begin{equation}
    t_{ab} = \bra{r_a^{\rho\sigma}}\tilde{T}_{\rho\sigma}\ket{l_b^{\rho\sigma}} \ ,
\end{equation}
where $\tilde{T}_{\rho\sigma}$ is the transfer matrix $T_{\rho\sigma}$ after the substitution $M^{(j)}\rightarrow\tilde{M}^{(j)}$.
In what follows, we assume a choice of $\tilde{M}^{(j)}$ such that $t_{00}\neq 0$. 
To compute the numerators in Eq.~\eqref{eqapp:norm2d_expanded} we first define the tensor
\begin{equation}
    \Theta^{ab}_{cd} = \partial_{\bar{M}^{(j)}}[\bar{T}_{\rho\sigma}]_{ab} \cdot \partial_{M^{(j)}}[T_{\rho\sigma}]_{cd} \ .
\end{equation}
By applying again the spectral decomposition, we obtain
\begin{equation}
\label{eqapp:TO_scalarprod}
\begin{split}
    \bar{C}_{\rho}^{(j)}\cdot C_{\rho}^{(j)} &= \sum_{a,a'=0}^{\chi^2-1} (\nu_a \bar{\nu}_{a'})^{N-1} \Theta^{aa}_{a'a'} \ , \\
    \bar{C}_{\rho}^{(j)}\cdot C_{\rho_{I_j}}^{(j)} &= \sum_{a,a',a''=0}^{\chi^2-1}\sum_{b,c=0}^{\chi-1} \bar{\nu}_{a'}^{N-1} \Theta^{aa''}_{a'a'} \nu_a^{\ell} \Gamma^{bc}_a (\lambda_b \mu_c)^{N-(2\ell+1)} \tilde{\Gamma}^{a''}_{bc} \nu_{a''}^{\ell} \ , \\
    \bar{C}_{\rho_{I_j}}^{(j)}\cdot C_{\rho_{I_j}}^{(j)} &= \sum_{a,a',a'',a'''=0}^{\chi^2-1}\sum_{b,b',c,c'=0}^{\chi-1} (\nu_a \bar{\nu}_{a'})^{\ell} \Gamma^{bc}_a \bar{\Gamma}^{b'c'}_{a'} (\lambda_b \bar{\lambda}_{b'} \mu_c \bar{\mu}_{c'})^{N-(2\ell+1)} \tilde{\Gamma}^{a''}_{bc} \bar{\tilde{\Gamma}}^{a'''}_{b'c'} (\nu_{a''} \bar{\nu}_{a'''})^{\ell} \Theta^{aa''}_{a'a'''} \ .
\end{split}
\end{equation}
At this point, in both the overlaps and the scalar products, we separate the term in the summation where all the $a$ indices are $0$ from the rest, i.e.
\begin{equation}
\label{eqapp:norm2_eps1}
    \frac{\bar{C}_{\rho}^{(j)}\cdot C_{\rho}^{(j)}}{\text{tr}[\rho\tilde{\sigma}]} = \frac{\sum_{a,a'} (\nu_a \bar{\nu}_{a'})^{N-1} \Theta^{aa}_{a'a'}}{\sum_{a,a'} (\nu_a \bar{\nu}_{a'})^{N-1}t_{aa}\bar{t}_{a'a'}} = \frac{(\nu_0 \bar{\nu}_{0'})^{N-1} \Theta^{00}_{00} + \sum_{a\neq 0 \lor a'\neq 0} (\nu_a \bar{\nu}_{a'})^{N-1} \Theta^{aa}_{a'a'}}{(\nu_0 \bar{\nu}_{0})^{N-1}|t_{00}|^2 + \sum_{a\neq 0\lor a'\neq 0} (\nu_a \bar{\nu}_{a'})^{N-1}t_{aa}\bar{t}_{a'a'}} = \frac{\Theta^{00}_{00}}{|t_{00}|^2}(1 + \varepsilon_1) \ ,
\end{equation}
where we have used that overlaps are real, hence $\text{tr}[\rho\tilde{\sigma}] = \sum_{a} t_{aa}\nu_a ^{N-1} = \sum_{a} \bar{t}_{aa} \bar{\nu}_a^{N-1}$, and $\varepsilon_1$ is
\begin{equation}
    \varepsilon_1 = \frac{\bar{C}_{\rho}^{(j)}\cdot C_{\rho}^{(j)}/(\text{tr}[\rho\tilde{\sigma}])^2}{(\bar{C}_{\rho}^{(j)}\cdot C_{\rho}^{(j)})_0/(\text{tr}[\rho\tilde{\sigma}]_0)^2} - 1 ~,~ (\bar{C}_{\rho}^{(j)}\cdot C_{\rho}^{(j)})_0 = (\nu_0 \bar{\nu}_{0})^{N-1} \Theta^{00}_{00} ~,~ \text{tr}[\rho\tilde{\sigma}]_0 = t_{00}\nu_0 ^{N-1} \ .
\end{equation}
By repeating this step with the two other terms in Eq.~\eqref{eqapp:norm2d_expanded}, we obtain the same form
\begin{equation}
\label{eqapp:epsilon_decomp}
    \frac{\bar{C}_{\rho}^{(j)}\cdot C_{\rho_{I_j}}^{(j)}}{\text{tr}[\rho\tilde{\sigma}] \text{tr}[\rho_{I_j}\tilde{\sigma}_{I_j}]} = \frac{\Theta^{00}_{00}}{|t_{00}|^2}(1 + \varepsilon_2),~
    \frac{\bar{C}_{\rho_{I_j}}^{(j)}\cdot C_{\rho_{I_j}}^{(j)}}{\text{tr}[\rho_{I_j}\tilde{\sigma}_{I_j}]^2} = \frac{\Theta^{00}_{00}}{|t_{00}|^2}(1 + \varepsilon_3) \ .
\end{equation}
Under the assumption that the spectrum of $T_{\rho\sigma}$ is gapped, this way of rewriting the terms in Eq.~\eqref{eqapp:norm2d_expanded} separates a leading contribution (given by the largest eigenvalues in the spectra of the transfer operators) from a sub-leading contribution.
Being the leading contribution constant, and equal among the three terms in Eq.~\eqref{eqapp:norm2d_expanded}, we obtain
\begin{equation}
\label{eqapp:norm2d_eps}
    \Bigg|\Bigg|\frac{C_{\rho}^{(j)}}{\text{tr}[\rho\tilde{\sigma}]} - \frac{C_{\rho_{I_j}}^{(j)}}{\text{tr}[\rho_{I_j}\tilde{\sigma}_{I_j}]}\Bigg|\Bigg|_2^2 = \frac{\Theta_{00}^{00}}{|t_{00}|^2}\left(\varepsilon_1 - 2 \text{Re}(\varepsilon_2) + \varepsilon_3\right) \ .
\end{equation}
Note that for $\Theta^{00}_{00} = 0$ the inequality Eq.~\eqref{eqapp:local_app_C_err} is trivial.
Therefore, in what follows we assume $\Theta^{00}_{00} \neq 0$.

\subsubsection{Upper bounds}

We now provide upper bounds for the terms $\{\varepsilon_m\}$.
To do so, we first define the ratios
\begin{equation}
    r_1 = \frac{\bar{C}_{\rho}^{(j)}\cdot C_{\rho}^{(j)}}{(\bar{C}_{\rho}^{(j)}\cdot C_{\rho}^{(j)})_0} - 1 ~,~ r_1' = \frac{\text{tr}[\rho\tilde{\sigma}]}{(\text{tr}[\rho\tilde{\sigma}])_0} - 1  \ .
\end{equation}
Similar ratios will appear in the computation of $\varepsilon_2$ and $\varepsilon_3$.
From the absolute values of these ratios, it is possible to derive upper bounds for $\varepsilon_1$ in the following way
\begin{equation}
\label{eqapp:eps1_bound1}
    |\varepsilon_1| = \Bigg|\frac{1+r_1}{(1+r_1')^2} - 1\Bigg|= \Bigg|\frac{r_1 - r_1'(2 + r_1')}{(1+r_1')^2}\Bigg| \leq \frac{|r_1| + |r_1'|(2 + |r_1'|)}{(1-|r_1'|)^2} \ . 
\end{equation}
Now we make use of the spectral decompositions of the transfer operators to derive upper bounds on the absolute values of $r_1$ and $r_1'$.
From the spectral decomposition in Eq.~\eqref{eqapp:TO_overlaps}, and from the definition of the correlation length $\xi_{\rho\sigma}$ we obtain
\begin{equation}
    |r_1'| \leq \sum_{a\neq0} |t_{aa}/t_{00}||\nu_a /\nu_0 |^{N-1} \leq \alpha\chi^2 |\nu_1 /\nu_0 |^{N-1} = \alpha\chi^2 e^{-(N-1) / \xi_{\rho\sigma}} \ ,
\end{equation}
where $\alpha = \text{max}_{ab}\left(|t_{ab}/t_{00}|\right)$.
To further simplify the upper bound Eq.~\eqref{eqapp:eps1_bound1}, we impose $|r_1'|\leq1/2$, which is satisfied if
\begin{equation}
    \alpha\chi^2 e^{-(N-1)/\xi_{\rho\sigma}} \leq 1/2 \rightarrow N \geq \xi_{\rho\sigma} \text{log}(2\alpha\chi^2)+1 \ . 
\end{equation}
Using the spectral decomposition Eq.~\eqref{eqapp:TO_scalarprod}, analogously to $r_1'$, for $r_1$ we have
\begin{equation}
\begin{split}
    |r_1| \leq \Bigg{|} \sum_{a\neq0 \lor a'\neq0} \left(\frac{\nu_a}{\nu_0} \frac{\bar{\nu}_{a'}}{\bar{\nu}_{0}}\right)^{N-1} \frac{\Theta^{aa}_{a'a'}}{\Theta^{00}_{00}} \Bigg{|} \leq 2\beta \chi^2 e^{-(N-1)/\xi_{\rho\sigma}} + \beta \chi^4 e^{-2(N-1)/\xi_{\rho\sigma}} \leq c_1 \chi^2 e^{-(N-1)/\xi_{\rho\sigma}} \ ,
\end{split}
\end{equation}
where $\beta = \text{max}_{abcd}\left(\frac{\Theta_{cd}^{ab}}{\Theta_{00}^{00}} \right) = O(1)$, and $c_1$ is a constant that depends on $\alpha$ and $\beta$.
Putting these results together, using Eq.~\eqref{eqapp:eps1_bound1} and $|r_1'|<1/2$, we finally obtain the upper bound for $|\varepsilon_1|$
\begin{equation}
\label{eqapp:eps1_bound2}
    |\varepsilon_1| \leq 4(|r_1| + 5|r_1'|/2) \leq K_1 \chi^2 e^{-(N-1)/\xi_{\rho\sigma}} \ ,
\end{equation}
where $K_1$ is a constant.

The upper bounds for $|\varepsilon_2|$ and $|\varepsilon_3|$ follow from similar steps.
In particular, for $|\varepsilon_2|$ we have the upper bound
\begin{equation}
\label{eqapp:eps2_bound1}
    |\varepsilon_2|\leq \Bigg| \frac{1 + r_2}{(1+r_1')(1+r_2')} - 1 \Bigg| \leq 2\frac{|r_2| + |r_1'| + 3|r_2'|/2}{1-|r_2'|} \ ,
\end{equation}
obtained in terms of the ratios
\begin{equation}
    r_2 = \frac{\bar{C}_{\rho}^{(j)}\cdot C_{\rho_{I_j}}^{(j)}}{(\bar{C}_{\rho}^{(j)}\cdot C_{\rho_{I_j}}^{(j)})_0} - 1 ~,~ r_2' = \frac{\text{tr}[\rho_{I_j}\sigma_{I_j}]}{(\text{tr}[\rho_{I_j}\sigma_{I_j}])_0} - 1  \ .
\end{equation}
First, from Eq.~\eqref{eqapp:TO_overlaps} and the definition above, we derive the upper bound for $|r_2'|$
\begin{equation}
    |r_2'| = \Bigg|\frac{\sum_{a\neq 0\lor b\neq 0}\sum_{c, d} t_{ab}\nu_a^{\ell}\nu_b^{\ell}\Gamma_b^{cd}(\lambda_c \mu_d)^{N-(2\ell+1)}\tilde{\Gamma}^a_{cd}}{\sum_{c,d}t_{00} \nu_0^{2\ell}\Gamma_0^{cd}(\lambda_c \mu_d)^{N-(2\ell+1)}\tilde{\Gamma}^0_{cd}}\Bigg| \leq \alpha\gamma\left( 2\sum_{a\neq 0} |\nu_a / \nu_0|^{\ell} + \left(\sum_{a\neq 0} |\nu_a / \nu_0|^{\ell}\right)^2 \right) \leq c_2'\chi^2 e^{-\ell/\xi_{\rho\sigma}}  \ ,
\end{equation}
where $\gamma = \text{max}_{abc}\left(\Bigg|\frac{\Gamma_a^{bc}\tilde{\Gamma}^a_{bc}}{\Gamma_0^{bc}\tilde{\Gamma}^0_{bc}}\Bigg|\right) = O(1) $, $c_2'$ is a constant that depends on $\alpha$ and $\gamma$, and we have assumed
\begin{equation}
    \ell \geq \xi_{\rho\sigma}\text{log}(2\chi^2) \ .
\end{equation}
Also in this case we assume $|r_2'|\leq 1/2$, which is satisfied when
\begin{equation}
    \ell\geq \xi_{\rho\sigma} \text{log}(2c_2'\chi^2) \ .
\end{equation}
From Eq.~\eqref{eqapp:TO_scalarprod}, proceeding in a similar way than for the ratios $r_1$ and $r_2'$, we obtain the upper bound for $|r_2|$
\begin{equation}
    |r_2| \leq \beta\gamma \left( 3\chi^2 e^{-\ell / \xi_{\rho\sigma}} + 3\chi^4 e^{-2\ell / \xi_{\rho\sigma}} + \chi^6 e^{-3\ell / \xi_{\rho\sigma}} \right) \leq c_2 \chi^2 e^{-\ell / \xi_{\rho\sigma}} \ ,
\end{equation}
where $c_2$ is a constant.
Combining the upper bounds for $|r_1'|$, $|r_2|$ and $|r_2'|$ in Eq.~\eqref{eqapp:eps2_bound1}, we obtain
\begin{equation}
\label{eqapp:eps2_bound2}    
    |\varepsilon_2| \leq 4(|r_2| + |r_1'| + 3|r_2'|/2) \leq K_2 \chi^2 e^{-\ell/\xi_{\rho\sigma}} \ ,
\end{equation}
where $K_2$ is a constant.
Without further assumptions, and repeating similar steps, we also obtain
\begin{equation}
\label{eqapp:eps3_bound}    
    |\varepsilon_3| \leq K_3 \chi^2 e^{-\ell/\xi_{\rho\sigma}} \ ,
\end{equation}
where $K_3$ is a constant.

\subsubsection{Final result}

As a final step, we combine the expression for the norm$-2$ distance in Eq.~\eqref{eqapp:norm2d_eps} with the upper bounds for the $\{|\varepsilon_m|\}$ to obtain Eq.~\eqref{eqapp:approx_update_mt}.
At this point, we can plug in this last equation the upper bounds for $\{|\varepsilon_m|\}$ Eqs.~\eqref{eqapp:eps1_bound2},\eqref{eqapp:eps2_bound2},\eqref{eqapp:eps3_bound}.
Under the assumptions that
\begin{enumerate}
    \item $\xi_{\rho\sigma}\geq0$,
    \item $\Gamma_0^{bc}\tilde{\Gamma}^0_{bc}\neq 0$, $t_{00}\neq 0$,
    \item $N \geq \xi_{\rho\sigma}\text{log}(2\alpha\chi^2) + 1$,
    \item $\ell \geq \xi_{\rho\sigma}\text{max}\left(\text{log}(2c_2'\chi^2), \text{log}(2\chi^2)\right)$ \ ,
\end{enumerate}
we obtain
\begin{equation}
    \Bigg|\Bigg|\frac{C_{\rho}^{(j)}}{\text{tr}[\rho\tilde{\sigma}]} - \frac{C_{\rho_{I_j}}^{(j)}}{\text{tr}[\rho_{I_j}\tilde{\sigma}_{I_j}]}\Bigg|\Bigg|_2^2 = \frac{\Theta_{00}^{00}}{|t_{00}|^2}\left(\varepsilon_1 - 2 \text{Re}(\varepsilon_2) + \varepsilon_3\right) \leq K \chi^2 e^{-\ell/\xi_{\rho\sigma}} \ ,
\end{equation}
where $K = \frac{\Theta_{00}^{00}}{|t_{00}|^2}\text{max}(K_1,K_2, K_3)$ is a constant that depends on $\alpha$, $\beta$ and $\gamma$.

Note that the same derivation can be now repeated by substituting $\rho\rightarrow\sigma$, obtaining

\section{Statistical errors in the learning algorithm}
\label{app:stat_err}

In this section we study analytically the effect of statistical errors in the learning algorithm due to a finite number of measurements $N_u^L N_M$.
We consider the framework of classical shadows obtained from local randomized measurements, which we briefly review in the first subsection.
Then, we show how to apply it to our learning algorithm, and prove an upper bound on the effect of statistical errors in the update of a single tensor $M^{(j)}$.
For a choice of the parameters $(\ell, \chi')$, we prove that the number of measurement bases needed to obtain the optimal tensor $M^{(j)}$ up to an error $\varepsilon$ scales like $N_u^L = O(2^{2\ell}{\chi'}\varepsilon^{-2})$, independently of the total system size $N$.

\subsection{Local randomized measurements and classical shadows}
\label{subapp:RM}

In this section we review the classical shadow framework for local randomized measurements, and highlight results which we use throughout the manuscript.
We refer the reader to Ref.~\cite{Elben_2022_review} for a more comprehensive review.

Let us assume we have prepared a $N-$qubit quantum state $\rho$ in an experiment.
Local randomized measurement consists in applying a layer of single-qubit random unitaries $\bigotimes_{j=1}^N u_j^{(r)}$, before measuring in the computational basis, where $r$ is the index associated with the bases, and each $u_i^{(r)}$ is sampled from the circular unitary ensemble (CUE).
We consider measurements in $N_u$ different random bases, with $N_M$ shots per basis, for a total of $M = N_u N_M$ experimental repetitions.
After each experimental shot, we obtain as a result a bitstring $s^{(r, m)}$, with $r\in[1, N_u]$, $m\in[1, N_M]$.
Such dataset can be then post-processed in various ways to extract properties of $\rho$.
Classical shadows represent a powerful method to post-process such data~\cite{Huang_2020_shadows}.
For an arbitrary subset of qubits $I$, we can define for the $r-$th measurement basis a classical shadow
\begin{equation}
    \hat{\rho}^{(r)}_I = \frac{1}{N_M}\sum_{m=1}^{N_M}\bigotimes_{j\in I}\left(3\left(u_j^{(r)}\right)^{\dagger}\ket{s_j^{(r, m)}}\bra{s_j^{(r, m)}}u_j^{(r)} - \mathbb{I}_j\right) \ ,
\end{equation}
where $\mathbb{I}_j$ is the identity on the $j-$th qubit.
Note that classical shadows only depend on the local unitaries $u_j^{(r)}$, which are classically sampled, and on the measurement outcomes $s^{(r,m)}_j$.
While for $N_M = 1$ classical shadows are product operators, and therefore numerically efficient, the sum over $N_M$ can be computed efficiently only for modest subsystem sizes $|I|$. 

The main property of classical shadows is that they are unbiased estimators of $\rho$, \textit{i.e.}
\begin{equation}
    \rho_I =  \mathbb{E}[\hat{\rho}_I^{(r)}] \ ,
\end{equation}
where $\mathbb{E}$ is the quantum mechanical average over all the possible bases and outcomes.
Using this property, it is possible to obtain an estimator for any linear function of $\rho$, e.g. for the expectation value of an observable $O$ with support $I$
\begin{equation}
\label{eqapp:estimator_observables}
    \hat{o} = \frac{1}{N_u}\sum_{r} \text{tr}[\hat{\rho}_I^{(r)} O] = \text{tr}[\hat{\rho}_I O] \ ,
\end{equation}
where we have defined the averaged shadow $\hat{\rho}_I = \frac{1}{N_u}\sum_{r}\hat{\rho}_I^{(r)}$.
The statistical errors of an estimator are related to its variance, which in this case, for $N_M = 1$, admits the upper bound~\cite{Huang_2020_shadows, Elben2020ptmoments}
\begin{equation}
\label{eqapp:var_observables}
    \text{Var}[\hat{o}] \leq \frac{2^{|I|}}{M}||O||_2^2 \ .
\end{equation}
Analogously, we can write estimators for multi-copy observables 
\begin{equation}
    \hat{o}^{(n)} = \frac{1}{n!}\binom{N_u}{n}^{-1}\sum_{r_1\neq r_2\neq...\neq r_n} \text{tr}[\hat{\rho}_I^{(r_1)}\hat{\rho}_I^{(r_2)}...\hat{\rho}_I^{(r_n)}O^{(n)}] 
\end{equation}
and compute associated bounds on their variance~\cite{rath2021qfi}, which generally scale exponentially with the support of $O^{(n)}$.

Relevant examples of such multi-copy observables include the moments of the reduced density matrix $\mathcal{P}_n(\rho_I) = \text{tr}[\rho_I^n]$.
However, it is also possible to estimate these moments in a different way, that does not rely on the knowledge of the unitaries $\{u_j^{(r)}\}$~\cite{Elben_2019_statistical, Brydges_2019}.
For example, the purity $\mathcal{P}_2(\rho_I)$ can be estimated as
\begin{equation}
\label{eqapp:brydges}
    \hat{\mathcal{P}}_2(\rho_I) = \frac{2^{|I|}}{N_u N_M (N_M - 1)}\sum_{r = 1}^{N_u}\sum_{m\neq m' =1}^{N_M} 2^{-D[s_I^{(r, m)}, s_I^{(r, m')}]} \ ,
\end{equation}
where $D[s, s'] = \sum_j |s_j-s_j'|$ is the Hamming distance between two bitstrings, and $s_I^{(r, m)}$ is the bitstring obtained from the $m-$th shot in the $r-$th basis restricted to the subsystem $I$.
This estimator also requires $M = O(2^{a|I|})$ measurements with $a = O(1)$ for finite accuracy~\cite{Elben_2019_statistical}, but is computationally less expensive than using classical shadows.
Throughout this work, we use Eq.~\eqref{eqapp:brydges} to estimate purities, and classical shadows otherwise.

\subsection{Statistical errors in the optimization of a single tensor}

In this section we study the statistical errors in the update of a local tensor using randomized measurement data.
The update rule Eq.~\eqref{eqapp:approx_update_mt} requires experimental data to estimate $C_{\rho_{I_j}}^{(j)} = \text{tr}[\rho_{I_j}\partial_{M^{(j)}}\sigma_{I_j}]$.
Being $C_{\rho_{I_j}}^{(j)}$ linear in $\rho$, we can use the estimator Eq.~\eqref{eqapp:estimator_observables} for each one of its terms, obtaining
\begin{equation}
\label{eqapp:estimator_gradient}
    \hat{C}_{\rho_{I_j}}^{(j)} = \frac{1}{N_u^L}\sum_{r=1}^{N_u^L}\text{tr}[\hat{\rho}_{I_j}^{(r)}\partial_{M^{(j)}}\sigma_{I_j}] \ ,
\end{equation}
where $\hat{\rho}_{I_j}^{(r)}$ are classical shadows on the subsystem $I_j = [j-\ell, j+\ell]$, and $N_u^L$ is the number of random measurement bases collected in the learning set. 
The statistical errors in this estimator propagate to the tensor $M^{(j)}$, which will be then updated $\hat{M}^{(j)} \propto \mathcal{C}_{\sigma_{I_j}}^{(j)-1}\hat{C}_{\rho_{I_j}}^{(j)}$.
Let us define $M'^{(j)} = \mathcal{C}_{\sigma_{I_j}}^{(j)-1}C_{\rho_{I_j}}^{(j)}\propto M^{(j)}$.
Using the Cauchy-Schwarz inequality, we obtain the propagation of errors
\begin{equation}
    ||\hat{M}'^{(j)} - M'^{(j)}||_2 \leq ||\mathcal{C}_{\sigma_{I_j}}^{(j)-1}||_{\infty} ||\hat{C}_{\rho_{I_j}}^{(j)} - C_{\rho_{I_j}} ^{(j)}||_2 \ ,
\end{equation}
where $||\cdot||_2$ is the Hilbert-Schmidt norm (see Eq.~\eqref{eqapp:vector_norm}), and $||O||_{\infty} = \text{max}_v\frac{||Ov||_2}{||v||_2}$ is the spectral norm.
We expect that the errors in the normalization of $M^{(j)}$ will amount to a constant factor.

We now study the statistical errors in norm$-2$ $||\hat{C}_{\rho_{I_j}}^{(j)} - C_{\rho_{I_j}}^{(j)}||_2$. 
The statistical errors of the estimator Eq.~\eqref{eqapp:estimator_gradient} admit an upper bound via the multidimensional Chebyshev inequality~\cite{AltomareCampiti+1994}
\begin{equation}
    \text{Pr}\left(|| \hat{V} - V||_2 \geq \varepsilon \right) \leq \frac{1}{\varepsilon^2} \mathbb{E}\left[||\hat{V} - V||_2^2\right] = \frac{1}{\varepsilon^2} \sum_j \text{Var}(\hat{V}_j) \ ,
\end{equation}
with $\mathbb{E}[\hat{V}] = V$, from which we obtain
\begin{equation}
\label{eqapp:chebyshev}
    \text{Pr}\left(|| \hat{C}_{\rho_{I_j}}^{(j)} - C_{\rho_{I_j}}^{(j)}||_2 \geq \varepsilon \right) \leq \frac{1}{\varepsilon^2} \sum_{s,s'}\sum_{l,r=1}^{\chi'} \text{Var}\left( [\hat{C}_{\rho_{I_j}}^{(j)}]_{l r s s'} \right) \ ,
\end{equation}
where we have assumed that the bond dimension of $\sigma$ is $\chi'$ for all links.
Assuming that there is $N_M = 1$ shot per random basis, we can apply the upper bound on the variance for estimators of expectation values Eq.~\eqref{eqapp:var_observables}, obtaining
\begin{equation}
\label{eqapp:var_singleterm}
    \text{Var}\left( [\hat{C}_{\rho_{I_j}}^{(j)}]_{l r s s'} \right) \leq \frac{2^{2\ell+1}}{N_u^L} \bra{rr}\mathcal{C}_{\sigma_{I_j}}^{(j)}\ket{ll} \ ,
\end{equation}
where $\ket{ll} = \ket{l}\otimes \ket{l}$, $\ket{rr} = \ket{r}\otimes \ket{r}$.
Summing over all the indices results in the upper bound
\begin{equation}
\label{eqapp:var_singleterm}
    \sum_{l, r, s, s'}\text{Var}\left( [\hat{C}_{\rho_{I_j}}^{(j)}]_{l r s s'} \right) \leq \frac{2^{2\ell+3}\chi'}{N_u^L} \bra{\delta}\mathcal{C}_{\sigma_{I_j}}^{(j)}\ket{\delta} \ ,
\end{equation}
where $\ket{\delta} = \frac{1}{\sqrt{\chi'}}\sum_{l}\ket{ll}$.
We expect that the term $\bra{\delta}\mathcal{C}_{\sigma_{I_j}}^{(j)}\ket{\delta}$ contributes with a factor proportional to the purity $\text{tr}[\sigma_{I_j}^2]$.
This can be seen for translation-invariant MPOs using the transfer operator formalism introduced in App.~\ref{app:proof_local_approx_update}.
At the leading order in the spectrum of $T^{(1)}_{\sigma}$ and $T^{(2)}_{\sigma}$, we obtain
\begin{equation}
    \bra{\delta}\mathcal{C}_{\sigma_{I_j}}^{(j)}\ket{\delta} \simeq \frac{\bra{r^{\sigma^2}_0}\delta\rangle \bra{\delta}l^{\sigma^2}_0\rangle}{\nu_0}\mu_0^{N-2\ell-1}\nu_0^{2\ell+1}\bra{r_0^{\sigma^2}}l_0^{\sigma}l_0^{\sigma}\rangle\bra{r_0^{\sigma}r_0^{\sigma}}l_0^{\sigma^2}\rangle \ ,
\end{equation}
while for the purity
\begin{equation}
    \text{tr}[\sigma_{I_j}^2] \simeq \mu_0^{N-2\ell-1}\nu_0^{2\ell+1}\bra{r_0^{\sigma^2}}l_0^{\sigma}l_0^{\sigma}\rangle\bra{r_0^{\sigma}r_0^{\sigma}}l_0^{\sigma^2}\rangle \ .
\end{equation}
This implies that, up to corrections of order $O(\chi'^2 e^{-\ell/\xi_{\sigma}^{(2)}})$, we have
\begin{equation}
    \bra{\delta}\mathcal{C}_{\sigma_{I_j}}^{(j)}\ket{\delta} \simeq c~\text{tr}[\sigma_{I_j}^2] \ ,
\end{equation}
where $c = \frac{\bra{r^{\sigma^2}_0}\delta\rangle \bra{\delta}l^{\sigma^2}_0\rangle}{\nu_0} = O(1)$ is a constant factor.
Under the assumption that $\ell>O\left(\xi_{\sigma}^{(2)}\text{log}(\chi'^2)\right)$ we can absorb the sub-leading terms in a constant $c'$, obtaining the final result
\begin{equation}
\label{eqapp:meas_budget_global}
    \text{Pr}\left(|| \hat{C}_{\rho_{I_j}}^{(j)} - C_{\rho_{I_j}}^{(j)}||_2 \geq \varepsilon \right) \leq c'\text{tr}[\sigma_{I_j}^2]\frac{2^{2\ell+3}\chi'}{\varepsilon^2 N_u^L} \ .
\end{equation}
This implies that the number of measurements needed to achieve a controlled statistical error in norm$-2$ $||\hat{M}^{(j)} - M^{(j)}||_2$ scales like $M^L = N_u^L N_M = O(2^{2\ell}{\chi'}\varepsilon^{-2})$, as reported in the main text, and is independent of the total system size $N$.

\section{Scalable estimation of mixed-state fidelities}
\label{app:AFC_fidelities}

In this section we discuss in more detail the AFC estimation of mixed-state fidelities introduced in the main text.
First, leveraging the formalism of transfer operators, we prove approximate factorization conditions (AFC) for the overlap of two TIMPDOs with finite correlation lengths.
Then, we argue that combining them with AFC for purities~\cite{vermersch2024AFC} allows us to define the AFC for the fidelities defined in the main text.
This implies that a measurement budget of size polynomial in $N$ is sufficient to estimate fidelities in these quantum states.
Lastly, we numerically demonstrate that a modest measurement budget is sufficient to estimate AFC fidelities in thermal states of up to $N = 128$ qubits.

\subsection{Mixed-state fidelities from randomized measurements}

First, we comment on the relation between the two mixed-state fidelities introduced in the main text: the GM fidelity (Eq.~\eqref{eqapp:GM_fidelity}) and the max fidelity, defined as~\cite{Liang_2019, Elben2020crossplatform}
\begin{equation}
    \mathcal{F}_{\text{max}} = \frac{\text{tr}[\rho\sigma]}{\text{max}(\text{tr}[\rho^2], \text{tr}[\sigma^2])} \ .
\end{equation}
We can see that $\mathcal{F}_{\text{GM}}(\rho, \sigma)\geq \mathcal{F}_{\text{max}}(\rho, \sigma)$, and $\mathcal{F}_{\text{max}}(\rho, \ket{\psi}\bra{\psi}) = \bra{\psi}\rho\ket{\psi}$, while the same does not hold for $\mathcal{F}_{\text{GM}}$.
Additionally, the GM fidelity is insensitive to certain noise models such as global depolarization, \textit{e.g.} $\mathcal{F}_{\text{GM}}(\rho, (1-p)\rho + p\mathbb{I}/2^N) \sim 1 - O(\text{tr}[\rho^2]/2^N) $.
This makes the max fidelity more suitable to physically distinguish quantum states, while the GM fidelity, being differentiable, is a better-suited cost function for optimization algorithms.

It is possible to evaluate both fidelities from randomized measurements by estimating sequentially the purities $\text{tr}[\rho^2]$ and $\text{tr}[\sigma^2]$ and their overlap $\text{tr}[\rho\sigma]$ (see Sec.~\ref{subapp:RM}).
In the context of our protocol, we are especially interested in estimating the fidelity between an experimentally prepared state $\rho$ and a known state $\sigma$; applying Eq.~\eqref{eqapp:var_observables}, we obtain 
\begin{equation}
    \text{Var}\left[\text{tr}[\hat{\rho}\sigma]\right] \leq \frac{2^N}{M}\text{tr}[\sigma^2] \ .
\end{equation}
This implies that the overlap can be estimated up to controlled statistical errors with $M = O(2^N)$ randomized measurements.
Given that estimating purities requires instead $M = O(4^N)$~\cite{rath2021qfi}, we generally need at least $M = O(4^N)$ measurements to estimate the previously-defined fidelities.

\subsection{Approximate factorization conditions for fidelities}

We now introduce approximate factorization conditions (AFC) for the above-defined fidelities, and show how they lift the requirement of exponentially-many randomized measurements for fidelity estimation.
First, let us recall AFC for purities, defined in Ref.~\cite{vermersch2024AFC}.
A quantum state $\rho$ satisfies AFC for purities if it is possible to approximate its global purity by
\begin{equation}
    \text{tr}[\rho^2] \simeq \mathcal{P}^{(k)}_{\text{AFC}}(\rho) = \frac{\text{tr}[\rho_{A_1 A_2}^2]\text{tr}[\rho_{A_2 A_3}^2]...\text{tr}[\rho_{A_{R-1} A_R}^2]}{\text{tr}[\rho_{A_2}^2]\text{tr}[\rho_{A_3}^2]...\text{tr}[\rho_{A_{R-1}}^2]} \ ,
\end{equation}
where $A_j$ are connected intervals of $|A_j| = k$ qubits, with relative error
\begin{equation}
    \Bigg|\frac{\mathcal{P}^{(k)}_{2,\text{AFC}}(\rho)}{\mathcal{P}_2(\rho)} -1\Bigg| \leq c_1\frac{N}{k}e^{-k/\xi} \ .
\end{equation}
For instance, a TIMPDO $\rho$ with bond dimension $\chi$ and correlation lengths $\xi_{\rho}^{(1)}, \xi_{\rho}^{(2)} = O(1)$ satisfies AFC for purities, with $\xi = \text{max}(\xi_{\rho}^{(1)}, \xi_{\rho}^{(2)})$, and $c_1 = O(\chi^2)$.
From this approximation, it has been proven that for $k>\xi$ the purity estimation is guaranteed to have a controlled error with $M = O(2^{4k}N^{3})$ measurements~\cite{vermersch2024AFC}.

In a similar fashion, we now define AFC for the overlap between two quantum states.
We say that two quantum states $\rho$ and $\sigma$ satisfy AFC for the overlap if the following approximation holds
\begin{equation}
    \text{tr}[\rho\sigma]\simeq\mathcal{O}^{(k)}_{\text{AFC}}(\rho, \sigma) = \frac{\text{tr}[\rho_{A_1 A_2}\sigma_{A_1 A_2}]\text{tr}[\rho_{A_2 A_3}\sigma_{A_2 A_3}]...\text{tr}[\rho_{A_{R-1} A_R}\sigma_{A_{R-1} A_R}]}{\text{tr}[\rho_{A_2}\sigma_{A_2}]\text{tr}[\rho_{A_3}\sigma_{A_3}]...\text{tr}[\rho_{A_{R-1}}\sigma_{A_{R-1}}]} \ ,
\end{equation}
with $A_j$ connected intervals of $|A_j| = k$ qubits, where the relative error is bounded by
\begin{equation}
\label{eqapp:AFC_overlap}
    \Bigg|\frac{\mathcal{O}^{(k)}_{\text{AFC}}(\rho, \sigma)}{\text{tr}[\rho\sigma]} -1\Bigg| \leq c_2 \frac{N}{k}e^{-k/\xi}
\end{equation}
for some constants $c_2$, $\xi$.
When two states $\rho$ and $\sigma$ satisfy both the AFC for the overlap and, respectively, for purities, we say that they satisfy AFC for the fidelity.
To justify this definition, consider a choice of $k$ for a pair of states $\rho$ and $\sigma$ satisfying AFC such that
\begin{equation}
    \Bigg|\frac{\mathcal{O}^{(k)}_{\text{AFC}}(\rho, \sigma)}{\mathcal{O}(\rho,\sigma)} -1\Bigg| \leq \varepsilon_{\mathcal{O}} ~,~ \Bigg|\frac{\mathcal{P}^{(k)}_{2,\text{AFC}}(\rho)}{\mathcal{P}_2(\rho)} -1\Bigg| \leq \varepsilon_{\mathcal{P}}' ~,~ \Bigg|\frac{\mathcal{P}^{(k)}_{2,\text{AFC}}(\sigma)}{\mathcal{P}_2(\sigma)} -1\Bigg| \leq \varepsilon_{\mathcal{P}}'' \ ,
\end{equation}
where $\varepsilon_{\mathcal{P}} = (\varepsilon_{\mathcal{P}}' +\varepsilon_{\mathcal{P}}''+\varepsilon_{\mathcal{P}}'\varepsilon_{\mathcal{P}}'')\leq 1/2$.
Then, it is possible to see that 
\begin{equation}
    \Bigg|\frac{\mathcal{F}^{(k)}_{\text{AFC}}(\rho, \sigma)}{\mathcal{F}(\rho,\sigma)} -1\Bigg| \leq 2(\varepsilon_{\mathcal{O}} + \varepsilon_{\mathcal{P}}) = O\left(\text{poly}\left(\frac{N}{k}e^{-k/\xi}\right)\right)
\end{equation}
for both $\mathcal{F} = \mathcal{F}_{\text{max}}$ and $\mathcal{F} = \mathcal{F}_{\text{GM}}$, where
\begin{equation}
    \mathcal{F}^{(k)}_{\text{max,AFC}}(\rho, \sigma) = \frac{\mathcal{O}^{(k)}_{\text{AFC}}(\rho, \sigma)}{\text{max}(\mathcal{P}^{(k)}_{2,\text{AFC}}(\rho), \mathcal{P}^{(k)}_{2,\text{AFC}}(\sigma))} ~,~ \mathcal{F}^{(k)}_{\text{GM,AFC}}(\rho, \sigma) = \frac{\mathcal{O}^{(k)}_{\text{AFC}}(\rho, \sigma)}{\sqrt{\mathcal{P}^{(k)}_{2,\text{AFC}}(\rho)\mathcal{P}^{(k)}_{2,\text{AFC}}(\sigma)}} \ .
\end{equation}
This means that, for a pair of states $\rho$ and $\sigma$ satisfying AFC for the fidelity, there exists a value of $k$ such that $\mathcal{F}^{(k)}_{\text{AFC}}(\rho, \sigma)\simeq\mathcal{F}(\rho,\sigma)$.
Note that this is trivially true for $\rho=\sigma$, as $\mathcal{F}^{(k)}_{\text{AFC}}(\rho, \rho)=1$ for any $k$. 
In App.~\ref{subapp:AFC_overlap} we prove that two translation-invariant MPDOs $\rho$ and $\sigma$ with finite correlation lengths satisfy the AFC for the overlap (hence for the fidelity), with $c_2 = O(\chi^2)$, $\chi$ being the largest bond dimension between $\rho$ and $\sigma$, and $\xi = \text{max}(\xi_{\rho}^{(1)}, \xi_{\sigma}^{(1)}, \xi_{\rho\sigma})$ (defined in App.~\ref{app:proof_local_approx_update}).

Finally, let us comment on the statistical errors in fidelity estimation using AFC.
In order to estimate the AFC purity of $N$ qubits with controlled errors $M =O(4^{2k}N^3)$ randomized measurements are sufficient~\cite{vermersch2024AFC}.
Repeating the same steps of the proof, it is possible to show that $M = O(2^{2k}N^3)$ randomized measurements are sufficient to estimate $\mathcal{O}^{(k)}_{\text{AFC}}(\rho, \sigma)$ with controlled errors.
This implies that, asymptotically, $O(4^{2k}N^3)$ measurements are sufficient to estimate AFC fidelities accurately, and we can in principle use this strategy to benchmark the results of the learning algorithm.
In App.~\ref{subapp:AFC_fidelity_numerics} we demonstrate numerically that the number of measurements needed is in fact within the reach of current experiments.

\subsection{Approximate factorization conditions for the overlap: proof for TIMPDOs}
\label{subapp:AFC_overlap}

In this section we prove that, under technical assumptions similar to the ones defined in Sec.~\ref{app:proof_local_approx_update}, two TIMPDOs satisfy AFC for the overlap.
In the language of Sec.~\ref{app:proof_local_approx_update}, we take two TIMPDOs $\rho$ and $\sigma$ such that
\begin{equation}
\label{eqapp:assumptions_overlap}
    \xi = \text{max}(\xi_{\rho}^{(1)}, \xi_{\sigma}^{(1)}, \xi_{\rho\sigma}) ~,~ \Gamma_0^{bc}\tilde{\Gamma}^0_{bc}\neq 0 \ .
\end{equation}
From the spectral decompositions of the transfer operators, we obtain the following expression for the overlap
\begin{equation}
    \mathcal{O}(\rho, \sigma) = \frac{\text{tr}[\rho\sigma]}{\text{tr}[\rho]\text{tr}[\sigma]} = \frac{\sum_{a=0}^{\chi^2-1} (\nu_a)^N}{\sum_{b,c=0}^{\chi-1}(\lambda_b \mu_{c})^N} = \left(\frac{\nu_0}{\lambda_0 \mu_0}\right)^N(1+\delta) \ ,
\end{equation}
where, from the definition
\begin{equation}
    \delta = \frac{(1 + \sum_{a=1}^{\chi^2-1}(\nu_a/\nu_0)^N) }{(1+\sum_{b=1}^{\chi-1}(\lambda_a/\lambda_0)^N)(1+\sum_{c=1}^{\chi-1}(\mu_a/\mu_0)^N)} - 1
\end{equation}
and the ones of the correlation lengths, we obtain the upper bound
\begin{equation}
\label{eqapp:bound_delta}
    |\delta| \leq 4(\chi^2e^{-N/\xi_{\rho\sigma}} + \chi e^{-N/\xi_{\rho}^{(1)}} + \chi e^{-N/\xi_{\sigma}^{(1)}} + \chi^2 e^{-2N/\text{max}(\xi_{\rho}^{(1)}, \xi_{\sigma}^{(1)})}) \leq 14\chi^2 e^{-N/\xi} \ ,
\end{equation}
with $\xi = \text{max}(\xi_{\rho}^{(1)}, \xi_{\sigma}^{(1)}, \xi_{\rho\sigma})$, and under the further assumptions that
\begin{equation}
    \Big|\sum_{b=1}^{\chi-1}(\lambda_a/\lambda_0)^N\Big|\leq 1/2 ~,~ \Big|\sum_{b=1}^{\chi-1}(\mu_a/\mu_0)^N\Big|\leq 1/2 \ ,
\end{equation}
which are satisfied for
\begin{equation}
    N \geq \xi~\text{log}(2\chi) \ .
\end{equation}
Following similar steps, we can also express the overlap on an interval $A_l$, with $|A_l| = k$, as
\begin{equation}
    \mathcal{O}(\rho_{A_l}, \sigma_{A_l}) = \text{tr}[\rho_{A_l}\sigma_{A_l}] = \sum_{abc}\nu_a^k \Gamma^{bc}_a(\lambda_b \mu_c)^{N-k}\tilde{\Gamma}^a_{bc} = \nu_0^k (\lambda_0 \mu_0)^{N-k} \Gamma^{00}_0 \tilde{\Gamma}_{00}^0(1 + \delta_k) \ .
\end{equation}
In this case, $\delta_k$ admits the upper bound
\begin{equation}
\label{eqapp:deltak_upper_bound}
    |\delta_k| \leq \beta(\chi^2 e^{-k/\xi} + 2\chi e^{-(N-k)/\xi} + 2\chi^3 e^{-N/\xi} + \chi^2 e^{-2(N-k)/\xi} + \chi^4 e^{-(2N-k)/\xi}) \leq \frac{7}{2}\beta \chi^2 e^{-k/ \xi} \ ,
\end{equation}
where $\beta = \text{max}_{abc}\left(\Bigg|\frac{\Gamma_a^{bc}\tilde{\Gamma}^a_{bc}}{\Gamma_0^{bc}\tilde{\Gamma}^0_{bc}}\Bigg|\right) $ is a constant, and we have assumed that
\begin{equation}
    \chi e^{-(N-2k)/\xi} \leq 1/2 \rightarrow N \geq 2k + \xi~\text{log}(2\chi) \ .
\end{equation}
Note that we can obtain a similar form for the overlap $\mathcal{O}(\rho_{A_lA_{l+1}}, \sigma_{A_lA_{l+1}})$ by simply performing a substitution $k\rightarrow 2k$.
From these two results, we can write the AFC overlap in periodic boundary conditions as
\begin{equation}
    \mathcal{O}^{(k)}_{\text{AFC}}(\rho, \sigma) = \frac{\prod_{l = 1}^R \mathcal{O}(\rho_{A_l A_{l+1}},\sigma_{A_l A_{l+1}})}{\prod_{l = 1}^R \mathcal{O}(\rho_{A_l}, \sigma_{A_l})} = \left(\frac{\nu_0}{\lambda_0 \mu_0}\right)^N \left(\frac{1+ \delta_{2k}}{1 + \delta_k}\right)^R = \left(\frac{\nu_0}{\lambda_0 \mu_0}\right)^N (1+\delta') \ ,
\end{equation}
where $R = N/k$, assuming $N$ is a multiple of $k$, and
\begin{equation}
    \delta' = \left(\frac{1+ \delta_{2k}}{1 + \delta_k}\right)^R - 1 \ .
\end{equation}
At this point, the errors induced by the AFC form can be expressed as
\begin{equation}
    \Bigg{|} \frac{\mathcal{O}^{(k)}_{\text{AFC}}(\rho, \sigma)}{\mathcal{O}(\rho, \sigma)} - 1 \Bigg{|} = \Bigg{|} \frac{1 + \delta'}{1 +\delta} - 1 \Bigg{|} \ .
\end{equation}
Now, let us assume that $|\delta|\leq 1/2$, which is satisfied when
\begin{equation}
    N \geq \xi~ \text{log}(28\chi^2) \ .
\end{equation}
Under this assumption, we obtain an upper bound on the error induced by the AFC form of the overlap as a function of $|\delta|$ and $|\delta'|$ 
\begin{equation}
\label{eqapp:bound_overlap}
    \Bigg{|} \frac{\mathcal{O}^{(k)}_{\text{AFC}}(\rho, \sigma)}{\mathcal{O}(\rho, \sigma)} - 1 \Bigg{|} \leq 2| \delta' - \delta | \leq 2(|\delta| + |\delta'|) \ .
\end{equation}
Since we already have an upper bound on $|\delta|$, the last step is to obtain an upper bound on $|\delta'|$.
First of all, let us define $\delta'' \geq \text{max}(|\delta_k|, |\delta_{2k}|)$, such that we can write
\begin{equation}
    |\delta'| = \Bigg{|} \left(\frac{1+\delta_{2k}}{1+\delta_k}\right)^R  - 1 \Bigg{|} \leq \Bigg{|} \left(\frac{1+\delta''}{1-\delta''}\right)^R - 1 \Bigg{|} \leq |e^{4R\delta''} - 1| \ .
\end{equation}
As a last assumption, we take
\begin{equation}
    4R\delta'' \leq 1  \ ,
\end{equation}
which is satisfied for
\begin{equation}
    k \geq  \xi~\text{log}(14\beta N\chi^2) \ .
\end{equation}
Under this condition, we can use that $e^{z}-1 \leq 2z$ for $0\leq z\leq 1$ and Eq.~\eqref{eqapp:deltak_upper_bound}, obtaining
\begin{equation}
    |\delta'| \leq |e^{4R\delta''} - 1| \leq 8R \delta'' \leq 28\beta \chi^2 \frac{N}{k}e^{-k/\xi} \ .
\end{equation}
At this point we can combine this bound with the one on $|\delta|$ (Eq.~\eqref{eqapp:bound_delta}) to obtain a bound on the AFC error for the overlap Eq.~\eqref{eqapp:bound_overlap}.
To summarize, assuming
\begin{enumerate}
    \item Eq.~\eqref{eqapp:assumptions_overlap}
    \item $N \geq \text{max}\left(2k + \xi~\text{log}(2\chi), \xi~\text{log}(28\chi^2)\right)$
    \item $k \geq \xi~\text{log}(14\beta N\chi^2)$ \ ,
\end{enumerate}
we finally obtain the upper bound on AFC errors in the overlap
\begin{equation}
    \Bigg{|} \frac{\mathcal{O}^{(k)}_{\text{AFC}}(\rho, \sigma)}{\mathcal{O}(\rho, \sigma)} - 1 \Bigg{|} \leq 28\chi^2 e^{-k/\xi}\left( e^{-(N-k)\xi} + 2\beta N / k\right) \leq 28\left( 2\beta+1\right)\chi^2\frac{N}{k} e^{-k/\xi} \ ,
\end{equation}
which corresponds to the approximate factorization conditions for the overlap Eq.~\eqref{eqapp:AFC_overlap}, with $c_2 = 28(2\beta + 1)\chi^2$ and $\xi = \text{max}(\xi_{\rho}^{(1)}, \xi_{\sigma}^{(1)}, \xi_{\rho\sigma})$, with $\beta = \text{max}_{abc}\left(\Bigg|\frac{\Gamma_a^{bc}\tilde{\Gamma}^a_{bc}}{\Gamma_0^{bc}\tilde{\Gamma}^0_{bc}}\Bigg|\right) $.

\subsection{Statistical errors in AFC fidelity estimation: numerical results}
\label{subapp:AFC_fidelity_numerics}

\begin{figure}[t]
    \centering
    \includegraphics[width=0.99\linewidth]{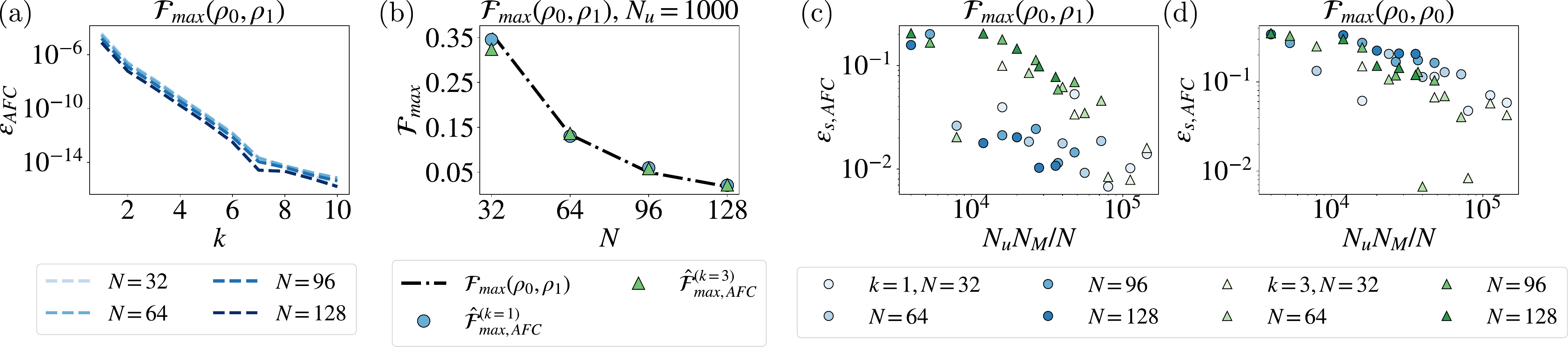}
    \caption{
    Numerical results for AFC fidelity estimation between the Ising-Gibbs states $\rho_0$ and $\rho_1$, defined in Eq.~\eqref{eqapp:Ising_Gibbs}.
    (a) Systematic relative error in the max fidelity $\mathcal{F}_{\text{max}}(\rho_0, \rho_1)$ due to the AFC assumption at finite $k$ for various $N$.
    (b) Fidelity estimation from randomized measurements, assuming AFC with $k = 1, 3$ of $\mathcal{F}_{\text{max}}(\rho_0, \rho_1)$ for various $N$ using $N_u = 1000$ random bases and $N_M = 512$.
    It is possible to see that the AFC estimation with a finite number of measurements (markers) faithfully captures the exact behavior of the fidelity (dotted line).
    (c,d) Relative errors between the AFC estimate and the exact value of fidelities for $k = 1,3$, $N_M = 512$, and various $N_u$ and $N$.
    In the regime considered, we observe a power-law decay of these errors as a function of $N_u N_M / N$.
    }
    \label{fig:S4}
\end{figure}

In this section, we numerically investigate the experimental feasibility of AFC fidelity estimation.
While we have previously argued that $O(4^{2k}\text{poly}(N))$ measurements are sufficient for an accurate estimation, this does not give a practical indication of the number of measurements typically needed.
Let us consider Ising-Gibbs states, defined as
\begin{equation}
\label{eqapp:Ising_Gibbs}
    \rho(\beta, g, h) = \frac{1}{Z}e^{-\beta H_{\text{Ising}}(g, h)} ~,~ H_{\text{Ising}}(g, h) = \frac{1}{4}\left(\sum_{j = 1}^{N-1}Z_j Z_{j+1} + \sum_{j=1}^N(gX_j + hZ_j)\right) \ ,
\end{equation}
where $\beta$ is the inverse temperature, and the factor $Z=\text{tr}[e^{-\beta H_{\text{Ising}}(g, h)}]$ ensures the normalization of the state.
These states are known to satisfy AFC for the purity at finite temperature~\cite{vermersch2024AFC, capel2024conditionalindependence}.
We consider the two Ising-Gibbs states $\rho_0 = \rho(\beta = 2, g = 1.01, h = 0.04)$ and $\rho_1 = (1, 1.5, 0)$ at sizes up to $N = 128$.
We numerically simulate these states as MPDOs using the ITensor library~\cite{fishman2022itensor} by evolving the infinite temperature state in imaginary time using time-evolving block decimation (TEBD).
We fix their bond dimension by fixing the truncation error to $10^{-20}$.
This choice results in bond dimension $\chi = (29, 20)$ for the states $(\rho_0, \rho_1)$ respectively.

As a first step, we study the relative error induced by the AFC approximation of the max fidelity $\mathcal{F}_{\text{max}}(\rho_0, \rho_1)$ between these two states at finite $k$, defined as
\begin{equation}
    \varepsilon_{\text{AFC}} = \Bigg|\frac{\mathcal{F}_{\max, \text{AFC}}^{(k)}(\rho_0, \rho_1)}{\mathcal{F}_{\max}(\rho_0, \rho_1)} - 1 \Bigg| \ ,
\end{equation}
and observe that it decays exponentially with $k$ (see Fig.~\ref{fig:S4}).
This is strong numerical evidence that $\rho_0$ and $\rho_1$ satisfy the AFC for the fidelity.
Then, using the RandomMeas library~\cite{RandomMeas}, we simulate local randomized measurements on $\rho_0$, and estimate its fidelity with both $\rho_1$ and itself as known target states.
We consider $N_M = 512$ shots per basis, and up to $N_u = 9000$ local random bases, and report the results in Fig.~\ref{fig:S4}.
First of all, it is possible to see that $N_u = 1000$ is already sufficient to accurately estimate these fidelities up to $N = 128$.
To perform a full scaling analysis we plot the relative error between the AFC estimation at finite $k = 1, 3$ from $M = N_u N_M$ measurements and the exact value of the fidelity, defined as
\begin{equation}
    \varepsilon_{s,\text{AFC}} = \Bigg|\frac{\hat{\mathcal{F}}^{(k)}_{\text{max,AFC}}(\rho_0, \rho_1)}{\mathcal{F}_{\text{max}}(\rho_0, \rho_1)} - 1 \Bigg| \ .
\end{equation}
First of all we observe that, in this regime, the statistical errors are orders of magnitude larger than the AFC errors.
This makes the AFC shadow estimator a good estimator of the fidelity, since its bias is negligible compared to its variance for an experimentally feasible number of measurements.
By considering various choices of $N$ and $N_u$, we observe a power-law decay of $\varepsilon_{s, \text{AFC}}$ with $N_uN_M / N$.
We note that the power-law decaying behavior of errors with the number of measurements is typical of randomized measurements, see for instance Ref.~\cite{Elben2020crossplatform}.
Furthermore, the data collapse hints at a linear scaling of the number of measurements required for an accurate estimation with respect to the total number of qubits $N$.

\section{Numerical results for the learning algorithm}
\label{app:numerical}

In this section we report on the performances of the learning algorithm by performing numerical simulations on Ising-Gibbs states, defined in Eq.~\eqref{eqapp:Ising_Gibbs}, with $(g, h) = (1.01, 0.04)$.
First, we study the performance of the learning algorithm as a function of $(\ell, \chi')$, with $N_u N_M \rightarrow \infty$.
In this case, the inputs of the learning algorithm are all the reduced density matrices on connected subsets of size $2(\ell+1)$ (we consider $2-$site updates).
This allows us to isolate the effect of the approximate local updates introduced in App.~\ref{app:algorithm}, which we show to be negligible at finite $\ell$ at various temperatures.
Finally, we perform a numerical experiment with finite $M^L = N_u^L N_M$ measurements at $\beta = 2$ to study the typical number of measurements required to learn such states.
This also serves to further validate the experimental results presented in the main text.

In both cases, we study the behavior of both the geometric-mean fidelity $\mathcal{F}_{\text{GM}}$ and the max fidelity $\mathcal{F}_{\text{max}}$.
While the former has a stronger dependence on the parameters of the learning algorithm, they generally display a similar behavior.  

\subsection{Approximate local updates}

\begin{figure}[t]
    \centering
    \includegraphics[width=0.99\linewidth]{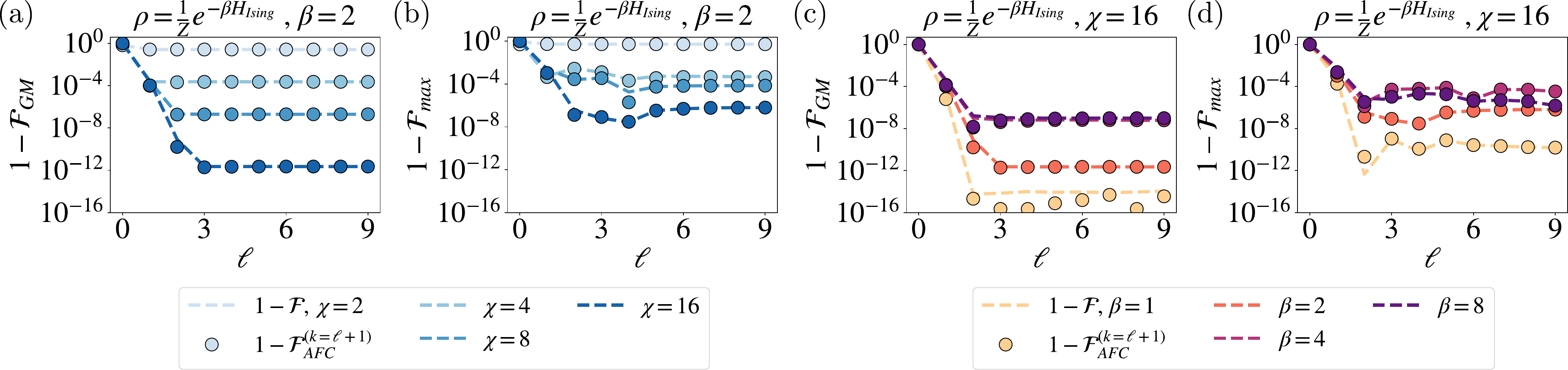}
    \caption{
    Numerical performances of the learning algorithm as a function of $(\ell, \chi')$ for $N_u^L \rightarrow\infty$.
    We consider Ising-Gibbs states as defined in Eq.~\eqref{eqapp:Ising_Gibbs} for $(g, h) = (1.01, 0.04)$ for various inverse temperatures $\beta$.
    We study both $\mathcal{F}_{\text{GM}}$ and $\mathcal{F}_{\text{max}}$, plotting both the global fidelities (dashed lines) and the AFC fidelities (markers) with $k = \ell + 1$.
    (a,b) Results for $\beta = 2$ as a function of $(\ell, \chi')$. We observe signatures of an exponential decay of the fidelity errors $1-\mathcal{F}$ with $\ell$, while the bond dimension $\chi'$ induces a lower bound on the fidelity error.
    The AFC fidelities generally match the global fidelities.
    (c, d) Results for $\chi' = 16$ as a function of $\ell$, for various inverse temperatures $\beta$.
    }
    \label{fig:S5}
\end{figure}

Here we numerically study the performance of the learning algorithm as a function of $(\ell, \chi^\prime)$ for $N_uN_M \rightarrow \infty$ on Ising-Gibbs states (Eq.~\eqref{eqapp:Ising_Gibbs}).
We generate MPOs approximating such states using the TEBD algorithm in imaginary time, implemented with the ITensor library~\cite{fishman2022itensor}.
By imposing a cut-off of $10^{-20}$ on singular values, we obtain MPOs with bond dimensions $\chi = (29, 45, 57)$ for $\beta = (2, 4, 8)$, and $N = 128$.
To mimic the experimental scenario where there is no notion of bond dimension, we learn these states up to a strictly smaller bond dimension $\chi^\prime < \chi$, where $\chi^\prime$ is the largest bond dimension allowed by the two-site updates.

We report the results in Fig.~\ref{fig:S5}.
In every plot, we show fidelity errors both computed exactly (dashed lines) and using AFC with $k = \ell + 1$ (markers).
The first one allows us to understand the behavior of the learning algorithm, while the second one is what we can reasonably access in an experimental scenario.
We generally observe a very good agreement between the two.

For $\beta = 2$ we observe an exponential decay of the fidelity errors $1- \mathcal{F}_{\text{GM}}$ and $1- \mathcal{F}_{\text{max}}$ as a function of $\ell$. 
Such decay saturates to a value dependent only on $\chi^\prime$, implying that only truncation errors are present at that point.
For $\ell = 1$ and $\chi^\prime = 4$ we already observe $1 - \mathcal{F}\sim 10^{-4}$ for both fidelities, meaning that for these parameters the state $\rho$ can be learned to a very high accuracy.
We observe a similar behavior for other values of $\beta$.
By fixing the bond dimension $\chi^\prime = 16$, we still observe an exponential decay of the fidelity errors with $\ell$, with smaller rates for higher $\beta$.

\subsection{Statistical errors}

\begin{figure}[t]
    \centering
    \includegraphics[width=0.75\linewidth]{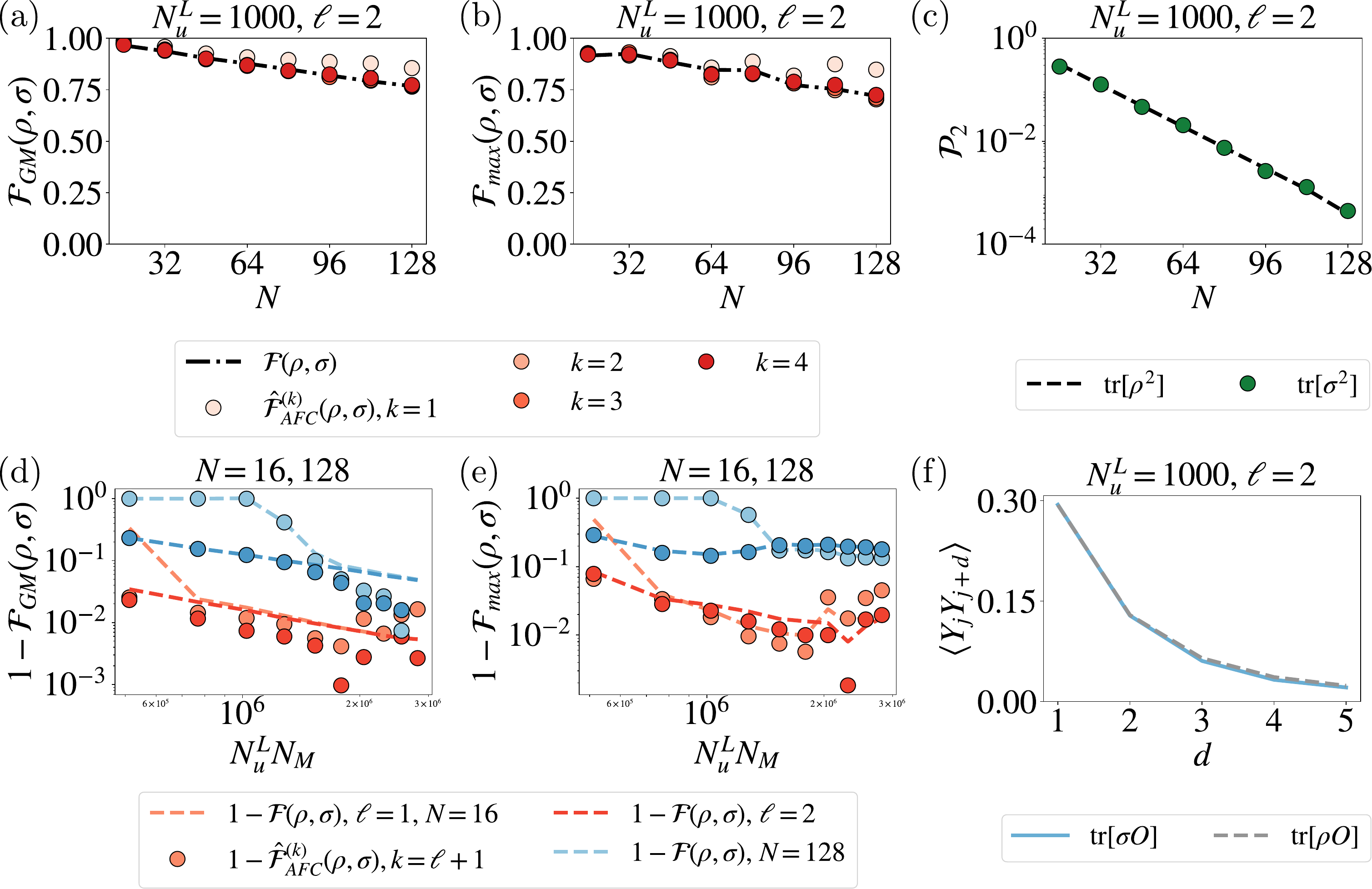}
    \caption{
    Numerical performances of the learning algorithm with a finite randomized measurement dataset.
    We consider Ising-Gibbs states (Eq.~\eqref{eqapp:Ising_Gibbs}) with $(\beta, g, h) = (2, 1.01, 0.04)$, and consider a dataset with $N_u^L +N_U^T = 10000$ bases, with $N_M = 1024$ shots per basis.
    (a,b) Fidelities as a function of the total number of qubits $N$ for $(\ell, \chi^\prime) = (2, 4)$, for which we know the learning algorithm outputs states with fidelities $\mathcal{F}\sim 1- 10^{-4}$.
    We plot both the exact fidelities (dotted lines) and the estimated AFC fidelity with the testing set for various $k$ (markers) for $N_u^L = 1000$, where only the latter are experimentally accessible.
    (c) Global purities of the target state $\rho$ and the output $\sigma$ for various $N$.
    (d,e) Fidelity errors as a function of $N_u^L$ for $N = 16, 128$, $\chi^\prime = 4$, and $\ell=1, 2$, both numerically exact (dotted lines) and from AFC estimation with $k = \ell + 1$ (markers).
    (f) Decay of two-point correlations as a function of the distance, compared between the target state (dashed line) and the learned MPO (straight line), averaged over $j\in [10, 113]$.
    }
    \label{fig:S6}
\end{figure}

In this section we study the effect of statistical errors on the learning algorithm, induced by a finite number of measurements $N_u N_M$.
To do so, we numerically sample randomized measurement outcomes from the MPO $\rho$, using the library RandomMeas~\cite{RandomMeas}.
Here, we consider $\rho$ to be the Ising-Gibbs state with $(\beta, g, h) = (2, 1.01, 0.04)$, which has bond dimension $\chi = 29$ for a cut-off $10^{-20}$ on singular values.
We collect data in a total of $N_u^L + N_u^T = 10^4$ bases, and consider $N_M = 1024$ shots per basis.
For each interval $I_j$ we construct the average shadow $\hat{\rho}_{I_j} = \frac{1}{N_u^L N_M}\sum_{r} \hat{\rho}_{I_j}^{(r)}$ from the randomized measurement data in the learning set, and use them as estimators of the reduced density matrices $\rho_{I_j}$ in the learning algorithm.
After each sweep we compute $\hat{\mathcal{F}}_{\text{max}}(\rho, \sigma)$ between the target state $\rho$ and the output $\sigma$ using the testing set, and AFC with $k = \ell +1$.
We then choose as final output the MPO $\sigma$ which maximizes this estimate of the fidelity.
In what follows we present the results of a single numerical experiment.
By repeating the numerical experiment, we do not observe significant differences in the results presented.

We run the learning algorithm for $\ell = 1, 2$ and $\chi^\prime = 4$, which we have seen to be sufficient to achieve a fidelity error $1-\mathcal{F}\sim 10^{-4}$.
We report the results in Fig.~\ref{fig:S6}.
For the parameters considered, we always observe fidelity errors $1-\mathcal{F}\gg 10^{-4}$, meaning that statistical errors are larger than the systematic errors due to finite $(\ell, \chi^\prime)$.
We study both $\mathcal{F}_{\text{GM}}$ and $\mathcal{F}_{\text{max}}$, and compute them exactly (dotted lines) and using the testing set under AFC with $k$ (markers).
These two estimates are in good agreement for $k>1$.

First, we study the fidelity errors for $N_u^L = 1000$ as a function of number of qubits $N$, from $N = 16$ to $N = 128$.
Similarly to the experiment reported in the main text, we observe a linear decay of both $\mathcal{F}_{\text{GM}}$ and $\mathcal{F}_{\text{max}}$, which hints at a linear propagation of statistical errors in the MPO (statistical errors in each tensor $M^{(j)}$ are independent of $N$, see App.~\ref{app:stat_err}).
For $(\ell, \chi^\prime) = (2, 4)$, $N_u^L = 1000$, we observe $\mathcal{F}_{\text{GM}} \sim 75\%$ and $\mathcal{F}_{\text{max}} \sim 70\%$.

For these parameters, we also compare physical properties between the MPOs $\rho$ and $\sigma$ for $N = 128$.
In particular, we consider the purity $\mathcal{P}_2(\rho) = \text{tr}[\rho^2]$, and the two-point correlation function $\langle Y_j Y_{j+d}\rangle$, where $Y_j$ is a local Pauli operator.
In both cases, we observe an almost perfect agreement between the predictions of $\rho$ and $\sigma$.

Finally, we study fidelity errors as a function of $N_u^L N_M$, with $N_u^L \in [1000, 5500]$, $N_M = 1024$.
We observe a power-law decay $1 - \mathcal{F}_{\text{GM}} \propto (N_u^L N_M)^{-0.9}$ in the limit of large $N_u^L N_M$.
While $\ell = 2$ already displays this behavior for $N_u^L = 1000$, the converse is not true for $\ell = 1$, which results in larger errors below a certain number of measurements.
We note a qualitatively similar behavior for $\mathcal{F}_{\text{max}}$, although its behavior is not strictly monotonous in this regime.

\section{Common randomized measurements}
\label{app:crm}

In this section we give details on the use of common randomized measurements (CRM)~\cite{vermersch2024common} in the experimental demonstration of the learning algorithm.
CRM allow one to reduce the overhead in number of bases $N_u$ when postprocessing randomized measurement data from a state $\rho$, given that a good classical approximation $\tau$ (also called prior) of $\rho$ is available.
In what follows, we review the key elements of common randomized measurements, and explain how to implement them in the learning algorithm.
Then, we describe the prior state used in the experimental demonstration, and compare performances with and without using CRM. 

\subsection{Common randomized measurements and tensor updates}

Let us consider an experimental quantum state $\rho$ on which we have collected randomized measurements, which we represent as a set of classical shadows $\{\hat{\rho}^{(r)}\}$.
In parallel, let us assume that we have access to a classical simulation of the experiment, which outputs the quantum state $\tau$.
CRM consist in defining a new set of unbiased estimators $\{\hat{\rho}_{\tau}^{(r)}\}$ such that, if $||\rho-\tau||_2\ll||\rho||_2$, the statistical errors in this new set of estimators is much smaller than conventional classical shadows.
From a classical representation of $\tau$, we can compute its classical shadows on a subsystem $A$ for $N_M \rightarrow\infty$ in the same bases in which we have performed randomized measurements on $\rho$, obtaining
\begin{equation}
    \hat{\tau}^{(r)}_A = \sum_{s, s'} 2^{-D[s,s']}P_{\tau_A, U}(s)\bigotimes_{j\in A} \left(u_j^{(r)\dagger}\ket{s'_j}\bra{s'_j}u_j^{(r)}\right) \ ,
\end{equation}
where $\ket{s} = \bigotimes_{j\in A}s_j$, $\ket{s'} = \bigotimes_{j\in A}s'_j$ are computational basis states on the subsystem $A$, $P_{\tau_A, U}(s) = \bra{s}U^{(r)\dagger}\tau_A U^{(r)}\ket{s}$ is their statistics for the state $\tau_A$ in a given random basis, with $U^{(r)} = \bigotimes_{j=1}^N u_j^{(r)}$, and $D[s,s']$ is the Hamming distance.
At this point, the new set of estimators is defined as
\begin{equation}
    \hat{\rho}_{\tau,A}^{(r)} = \hat{\rho}_A^{(r)} - \hat{\tau}_A^{(r)} + \tau_A \ .
\end{equation}
Note that these new estimators are unbiased, as $\mathbb{E}[\hat{\rho}^{(r)}_{\tau, A}] = \mathbb{E}[\hat{\rho}_A^{(r)}] = \rho_A$, and $\hat{\rho}^{(r)}_{\tau,A} = \rho_A$ in the extreme case where $\rho = \tau$ and $N_M\rightarrow\infty$.
To reach the latter regime, it is sufficient to take at least $N_M \gtrsim 2^{|A|}$ shots per basis, such that we obtain a good inference on the distribution $P_{\rho_A, U}(s)$.
More concretely, when estimating the expectation value of a Pauli observable with support on $A$ using measurements in $N_u$ random Pauli bases taking $N_M$ shots per basis, we obtain the variance~\cite{vermersch2024common}
\begin{equation}
\label{eqapp:CRM_obs_bound}
    \text{Var}[\text{tr}[\hat{\rho}_{\tau,A}O]] \leq \frac{3^{|A|}}{N_u}\left(\text{tr}[(\rho-\tau)O]^2 + \frac{1}{N_M}\right)   \ ,
\end{equation}
where $\hat{\rho}_{\tau,A} = \frac{1}{N_u}\sum_{r}(\hat{\rho}_A^{(r)} - \hat{\tau}_A^{(r)} + \tau_A)$, which results in a more precise estimation with respect to conventional classical shadows when $|\text{tr}[\rho_A O]|<|\text{tr}[(\rho_A-\tau_A) O]|$, given that $N_M$ is large enough.

We implement CRM in the learning algorithm in a similar way.
First, we perform a MPO simulation of the reduced density matrices $\tau_{I_j}$ based on a prior knowledge of $\rho$, such as the quantum circuit preparing $\rho$ plus some details about the noise, or the Hamiltonian of the system when considering systems at finite temperature.
Then, we modify the update rule in Eq.~\eqref{eqapp:approx_update_mt} as
\begin{equation}
\label{eqapp:CRM_update}
    M^{(j)}\leftarrow \mathcal{C}^{(j)-1}_{\sigma_{I_j}} \left(\hat{C}^{(j)}_{\rho_{I_j}} - \hat{C}^{(j)}_{\tau_{I_j}} + C^{(j)}_{\tau_{I_j}} \right) ~,~ M^{(j)}\leftarrow\frac{1}{\text{tr}[\sigma]}M^{(j)} \ ,
\end{equation}
where
\begin{equation}
\label{eqapp:CRM_gradientterm}
    \hat{C}^{(j)}_{\rho_{I_j}} - \hat{C}^{(j)}_{\tau_{I_j}} + C^{(j)}_{\tau_{I_j}} = \text{tr}[(\hat{\rho}_{I_j} - \hat{\tau}_{I_j} + \tau)\partial_{M^{(j)}}\sigma]
\end{equation}
is the CRM estimator of $C^{(j)}_{\rho_{I_j}}$.
While the analytical bound Eq.~\eqref{eqapp:CRM_obs_bound} only works for Pauli observables, we expect the CRM tensor update Eq.~\eqref{eqapp:CRM_update} to be less sensitive to statistical errors when $\tau\sim\rho$.
We study its performances empirically in the next section on experimental data.

\subsection{Experimental results}

\begin{figure}[t]
    \centering
    \includegraphics[width=0.7\linewidth]{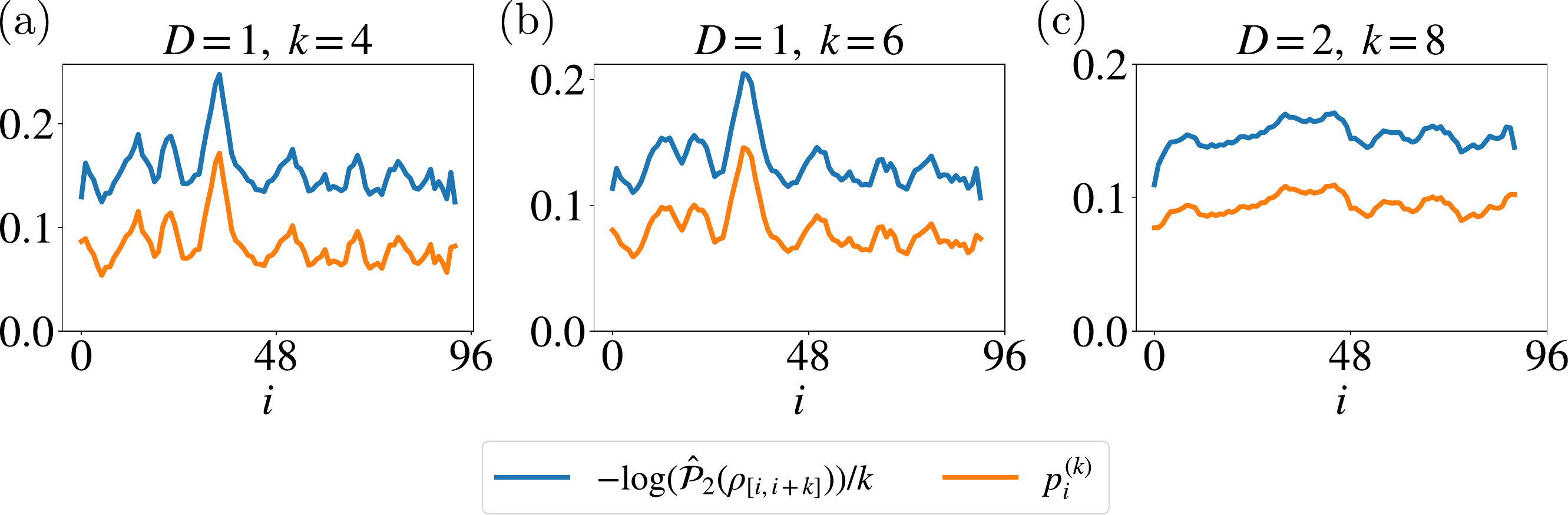}
    \caption{
    Estimated R\'enyi entropies of subsystems of size $k$ on quantum simulation of the kicked Ising model at depth $D$ versus the resulting local depolarization amplitude on the subsystem.
    We report results for the cases corresponding to $\ell = 1, 2$ in the learning algorithm for $D = 1$, and for $\ell = 3$ for $D = 2$.
    }
    \label{fig:S7}
\end{figure}

We start by describing the prior states $\{\tau_{I_j}\}$ used for CRM estimation from experimental data in the main text.
We recall that the quantum circuit implemented in the experiment aims to realize the pure state
\begin{equation}
    \ket{\psi} = \left(\prod_{j=1}^{N_{\text{tot}}-1}e^{i\frac{\pi}{4} Z_jZ_{j+1}} \prod_{j=1}^{N_{\text{tot}}}e^{-i\frac{\pi}{8} X_j}\right)^D \ket{0}^{\otimes N} \ ,
\end{equation}
where $D=1, 2$ is the depth of the circuit, $X_j$ and $Z_j$ are local Pauli operators.
The effect of experimental noise effectively result in the preparation of an unknown mixed state $\rho$.
For $D = 1, 2$ such states are exact MPS with bond dimension $\chi = 2, 4$.
$\tau = \ket{\psi}\bra{\psi}$ is not a useful prior state, since the condition $||\rho-\tau||_2\ll ||\rho||_2$ cannot be satisfied for the values of purity $\mathcal{P}_2(\rho)$ observed.
In particular, the norm$-2$ distance is lower bounded by the difference of the purities, i.e. $||\rho-\ket{\psi}\bra{\psi}||_2 \geq |1 - \mathcal{P}_2(\rho)| \simeq 1 - 10^{-4}$ for $D = 1$.

We take instead reduced density matrices $\tau_{I_j}' = \text{tr}_{\bar{I}_j}[\ket{\psi}\bra{\psi}]$, and add space-dependent noise using information collected from the experiment.
For each reduced density matrix $\tau_{[i, i+k]}'$, we obtain the prior $\tau_{[i, i+k]}$ as
\begin{equation}
    \tau_{[i, i+k]} = \left(\prod_{j\in[i,i+k]}\mathcal{E}_j(p_i^{(k)})\right) [\tau_{[i, i+k]}] \ ,
\end{equation}
where each $\mathcal{E}_j(p)[\rho] = (1-p)\rho + p~\text{tr}_j[\rho]\otimes \mathbb{I}_j$ is a local depolarizing channel with strength $p$.
For each interval $[i,i+k]$ we consider a homogeneous depolarizing strength $p_i^{(k)}$, which we fit by comparing the resulting purity $\text{tr}[\tau_{[i, i+k]}^2]$ to $\text{tr}[\rho_{[i,i+k]}^2]$ as estimated from randomized measurement data.
In this way we obtain a coarse-grained noise model for $\tau$.
We plot in Fig.~\ref{fig:S7} the observed purities and the corresponding results for the fitting for $D = 1, 2$ and $k = 2(\ell + 1)$, with the values of $\ell$ considered in the main text.

\begin{figure}[t]
    \centering
    \includegraphics[width=0.7\linewidth]{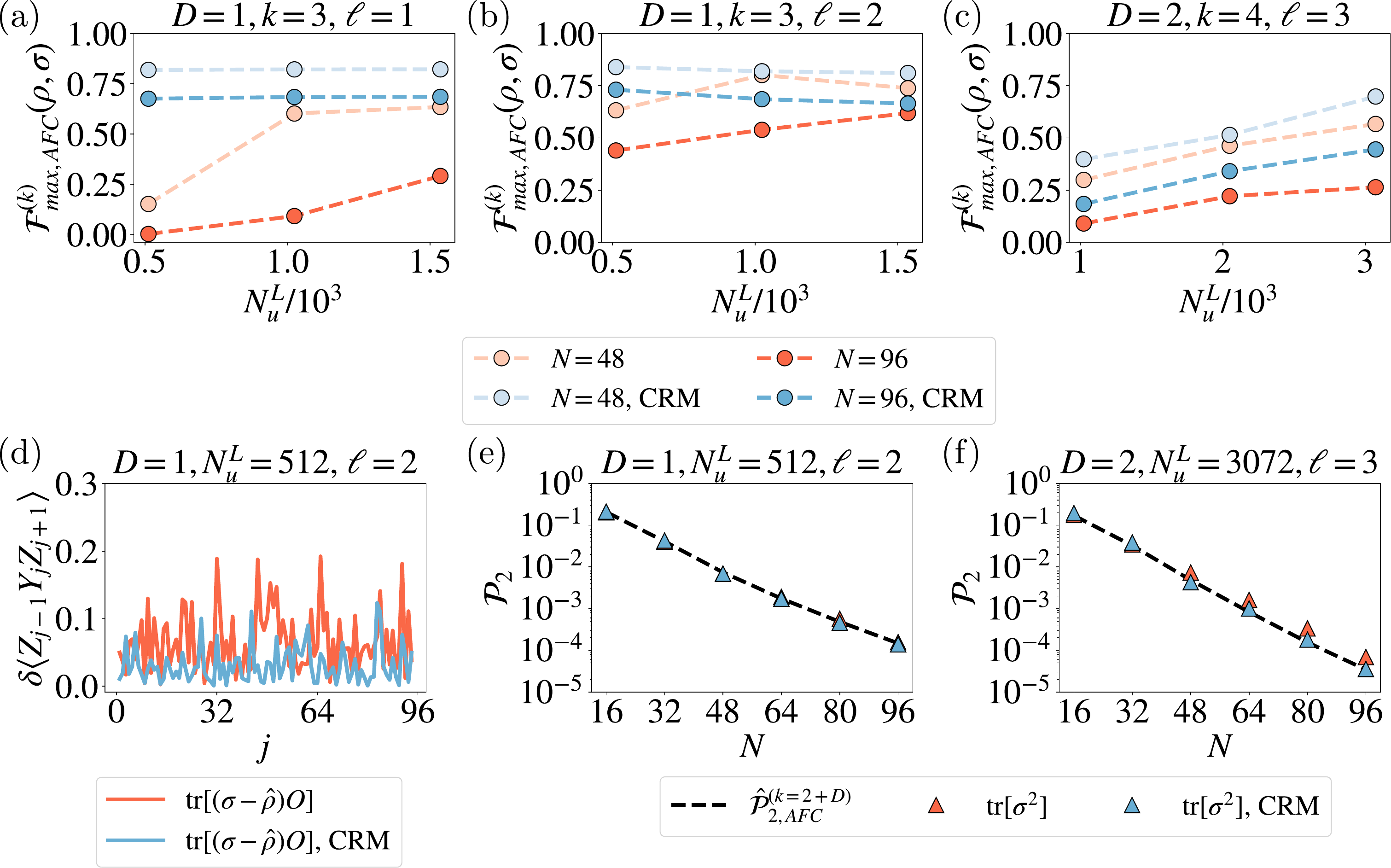}
    \caption{
    Performances of the learning algorithm with the experimental data with and without common randomized measurements.
    (a-c) Estimated AFC fidelity of the output as a function of the number of random bases in the learning set $N_u^L$ for $N = 48, 96$ with (blue lines) and without (red lines) common randomized measurements.
    We consider $\ell = 1, 2$ and $\chi' = 4$ for $D = 1$, $\chi' = 8$ and $\ell = 3$ for $D = 2$, considering AFC fidelities with parameter $k = 3$ for $D = 1$ and $k = 4$ for $D = 2$.
    (d) Difference in the estimation of the expectation value $\langle Z_{j-1}X_j Z_{j+1}\rangle$ between randomized measurements and the MPO learned from the full quantum state ($N = 96$ qubits).
    We plot the result both using common randomized measurements in the learning algorithm (blue lines) and not using them (red lines).
    (e,f) Global purities as a function of $N$ for $D = 1, 2$, comparing the AFC estimation (dashed lines) for $k = 3$ ($k = 4$) for $D = 1$ ($D = 2$) with the results of the learning algorithm for $(\ell, \chi') = (2, 4)$ ($(\ell, \chi') = (3, 8)$) with (blue markers) and without common randomized measurements (red markers). 
    }
    \label{fig:S8}
\end{figure}

In Fig.~\ref{fig:S8} we compare the results of the learning algorithm when analyzing the experimental data with and without CRM.
In all the cases we investigate, we observe that CRM equipped with the noise model described above allow to reach higher fidelities, as expected.
However, apart from the case $D =1$, $\ell = 1$, the increase in fidelity is modest, \textit{i.e.} comparable with roughly doubling $N_u^L$.
A similar consideration can be drawn for purities, for which we observe similar predictions with and without CRM for the cases considered in the main text.
Finally, we study the errors in estimating observables between our learning protocol and conventional randomized measurements.
We define the error in estimating an observable using the learning algorithm as
\begin{equation}
    \delta\langle O \rangle = |\text{tr}[\hat{\rho}O]-\text{tr}[\sigma O]| \ ,
\end{equation}
where $\hat{\rho} = \frac{1}{N_u^T N_M}\sum_{r=1}^{N_u^T N_M}\hat{\rho}^{(r)}$ is the average classical shadow obtained from the testing set, and $\sigma$ is the MPO output by the learning algorithm.
For $N_u^L = 512$, $N_u^T = 1536$, we observe that for $D=1$, $\ell = 2$ the errors in estimating $\langle X_{j-1} Z_j X_{j+1}\rangle$ are generally mitigated, with the largest error as a function of $j$ being reduced by half.
These results show how CRM can be useful to improve the predictive power of our learning algorithm.
Finally, we remark that a different prior $\tau$ can give drastically different results.
For instance, we expect that using a more fine-grained noise model (such as inhomogeneous local depolarization) would result in a better improvement of our results, at the cost of increased computational overhead to fit the model.

\end{document}